\documentclass[reprint,superscriptaddress,amsmath,amssymb,aps,prd,twocolumn,floatfix,nofootinbib]{revtex4-1}
\usepackage{amsmath, amssymb}
\usepackage{listings}
\usepackage{graphicx}
\usepackage{dcolumn}
\usepackage{bm}
\usepackage{float,color}
\usepackage{xcolor}
\usepackage[normalem]{ulem}
\usepackage{tabularx}
\usepackage{mathtools} 
\usepackage{multirow}
\usepackage{amssymb}
\usepackage{scrextend}
\usepackage{hyperref}
\usepackage{tikz}
\usepackage{enumitem}

\definecolor{ocre}{RGB}{52,177,201}

\begin{document}
\title{
Confronting general relativity with principal component analysis: Simulations and results from GWTC-3 events}
\author{Parthapratim Mahapatra}\email{MahapatraP@cardiff.ac.uk}
\affiliation{School of Physics and Astronomy, Cardiff University, Cardiff, CF24 3AA, United Kingdom}
\affiliation{Chennai Mathematical Institute, Siruseri, 603103, India}
\author{Sayantani Datta}
\affiliation{Chennai Mathematical Institute, Siruseri, 603103, India}
\affiliation{Department of Physics, University of Virginia, P.O. Box 400714, Charlottesville, Virginia 22904-4714, USA}
\author{Ish Gupta}
\affiliation{Institute for Gravitation and the Cosmos, Department of Physics, Penn State University, University Park, Pennsylvania 16802, USA}
\author{Poulami Dutta Roy}
\affiliation{Chennai Mathematical Institute, Siruseri, 603103, India}
\affiliation{Department of Physics, University of Florida, PO Box 118440, Gainesville, Florida 32611-8440, USA}
\author{Muhammed Saleem}
\affiliation{Center for Gravitational Physics, The University of Texas at Austin, Austin, Texas 78712, USA}
\author{Purnima Narayan}
\affiliation{Department of Physics and Astronomy, The University of Mississippi, University, Mississippi 38677, USA}
\author{Soumen Roy}
\affiliation{Centre for Cosmology, Particle Physics and Phenomenology - CP3, Université Catholique de Louvain, Louvain-La-Neuve, B-1348, Belgium}
\affiliation{Royal Observatory of Belgium, Avenue Circulaire, 3, 1180 Uccle, Belgium}
\author{Jan Steinhoff}
\affiliation{Max-Planck-Institute for Gravitational Physics (Albert-Einstein-Institute),
Am Mühlenberg 1, 14476 Potsdam-Golm, Germany}
\author{Deirdre Shoemaker}
\affiliation{Center for Gravitational Physics, The University of Texas at Austin, Austin, Texas 78712, USA}
\author{Alan J.~Weinstein}
\affiliation{LIGO Laboratory, California Institute of Technology, Pasadena, California 91125, USA}
\author{Anuradha Gupta}
\affiliation{Department of Physics and Astronomy, The University of Mississippi, University, Mississippi 38677, USA}
\author{B.~S.~Sathyaprakash}
\affiliation{Institute for Gravitation and the Cosmos, Department of Physics, Penn State University, University Park, Pennsylvania 16802, USA}
\affiliation{Department of Astronomy and Astrophysics, Penn State University, University Park, Pennsylvania 16802, USA}
\affiliation{School of Physics and Astronomy, Cardiff University, Cardiff, CF24 3AA, United Kingdom}
\author{K.~G.~Arun}
\affiliation{Chennai Mathematical Institute, Siruseri, 603103, India}
\affiliation{Institute for Gravitation and the Cosmos, Department of Physics, Penn State University, University Park, Pennsylvania 16802, USA}
\date{\today}

\begin{abstract}
We present a comprehensive assessment of multiparameter tests of general relativity (GR) in the inspiral regime of compact binary coalescences using principal component analysis (PCA). Our analysis is based on an extensive set of simulated gravitational-wave (GW) signals, including both general relativistic and non-GR sources, injected into zero-noise data colored by the noise power spectral densities of the LIGO and Virgo GW detectors at their designed sensitivities. We evaluate the performance of PCA-based methods in the context of two established frameworks: {\tt TIGER} and {\tt FTI}. For GR-consistent signals, we find that PCA enables stringent constraints on potential deviations from GR, even in the presence of multiple free parameters. Applying the method to simulated signals that explicitly violate GR, we demonstrate that PCA is effective at identifying such deviations. We further test the method using numerical relativity waveforms of eccentric binary black hole systems and show that missing physical effects—such as orbital eccentricity—can lead to apparent violations of GR if not properly included in the waveform models used for analysis.
Finally, we apply our PCA-based test to selected real gravitational-wave events from GWTC-3, including GW190814 and GW190412. We present joint constraints from selected binary black hole events in GWTC-3, finding that the 90\% credible bound on the most informative PCA parameter is $0.03^{+0.08}_{-0.08}$ in the {\tt TIGER} framework and  $-0.01^{+0.05}_{-0.04}$ in the {\tt FTI} framework, both of which are consistent with GR.
These results highlight the sensitivity and robustness of the PCA-based approach and demonstrate its readiness for application to future observational data from the fourth observing runs of LIGO, Virgo, and KAGRA.
\end{abstract}

\maketitle
\section{Introduction}
\label{sec:intro}
Parametric models play a central role in searches for new physics beyond the established paradigms, such as potential deviations from general relativity (GR)~\cite{GW150914-TGR,O2-TGR,GWTC2-TGR,GWTC3-TGR,GW170817-TGR,YYP2016}. 
Since these models aim to capture hitherto unknown physics, they often involve a high-dimensional parameter space.  
When tested against data, this high dimensionality can hinder inference by diluting constraints or inflating uncertainties. Principal component analysis (PCA), a linear algebra technique, offers an effective solution to this challenge. PCA reduces the effective dimensionality of the parameter space by identifying the best-measured linear combinations of the original parameter set without compromising on the information content in the data. It has been used recently in the context of gravitational-wave-based (GW-based) tests of GR~\cite{AP12, APrev13,ShoomEtalSVD,PCA1,Volkel:2022aca,PCA2,PCA3}. 

Our focus here is on parametrized tests of compact binary inspirals~\cite{BSat94,BSat95,AIQS06a,AIQS06b,YunesPretorius09, MAIS10,Li:2011cg,TIGER,Meidam:2017dgf,FTI}. The adiabatic inspiral phase of a compact binary is well described by post-Newtonian (PN) theory (see Ref.~\cite{Bliving} for a review), in which the amplitude and phase of the gravitational-wave signal are expanded in powers of the characteristic orbital velocity $v$ of the binary~\cite{BDI95}. Since the phase evolution carries more critical information for parameter inference than the amplitude, a widely used approach for parametric models of the GW phase in a non-GR theory is to introduce additive deviations to the PN phasing coefficients at various orders---eight in total till 3.5PN corresponding to ${\cal O}(v/c)^7$~\cite{MAIS10,TIGER,FTI}. Recently, Ref.~\cite{Blanchet:2023bwj} reported the 4.5PN phasing expression for quasicircular, nonspinning compact binaries; however, in this work, we restrict ourselves to the 3.5PN order since the newly computed terms are yet to be integrated into the standard tests of GR. 

The PN phasing coefficients elegantly capture the richness of nonlinear effects in the compact binary dynamics (such as ‘‘tail’’ effects \cite{BD88,BS93,BDI95}, tails of tails~\cite{B98tail}, tail square~\cite{B98quad}, memory effects \cite{Chr91,Th92,ABIQ04,Favata08}, spin-orbit effects \cite{KWWi93, Blanchet:2006gy}, spin-spin effects \cite{KWWi93,BFH2012}, etc.) at different PN orders in GR. Since these effects in a non-GR theory can differ from GR (see, for instance, Ref.~\cite{Bernard:2022noq} for scalar-tensor gravity theories), the additive parameters in the phasing are sensitive to such deviations. The parametrization is termed \emph{null} because all deviations vanish identically in GR; thus, any statistically significant departure from the null hypothesis could hint at physics beyond GR. This formulation of the inspiral phase enables several distinct strategies for testing GR, which we now discuss.

\subsection{Single versus multiparameter tests}
\label{sec:single_vs_multi}
There are different ways in which these tests can be performed.~\footnote{
Currently, the inspiral tests performed in the LVK papers use ten PN deformation parameters. These include eight fractional PN deformation parameters corresponding to the eight non-vanishing PN coefficients in GR up to 3.5PN order, along with absolute PN deformation parameters at the $-1$PN dipole term and the 0.5PN term. We do not include the absolute deformation terms in this analysis and consider deformations only to those which are nonzero in GR.} 
Along with the usual GR parameters, one can choose to (a) estimate only one parameter at a time (keeping all other testing GR parameters fixed at zero), (b) estimate all the parameters simultaneously, or (c) estimate a subset of them together while keeping the others fixed at zero. The single-parameter tests in (a) are the simplest among the tests that can be performed and are the ones implemented by the LIGO-Virgo-KAGRA (LVK) Collaboration~\cite{GW150914-TGR,O2-TGR,GW170817-TGR,GWTC3-TGR}, but they are subject to a fundamental limitation related to the theoretical framework of gravity. Should the true theory of gravity differ from GR, several phasing coefficients may deviate from their GR values. Therefore, a test of GR where all eight (or a sufficiently large number of) PN deformation coefficients are varied simultaneously would be ideal. Such a test involves estimating several additional parameters besides the standard GR ones from the data. This is a daunting task due to the strong degeneracies between the parameters that lead to uninformative posteriors. For example, Ref.~\cite{GW150914-TGR} performed such a multiparameter test on GW150914~\cite{GW150914} and found uninformative posteriors on the PN deformation parameters due to the aforementioned correlations among parameters (see Fig.~7 of Ref.~\cite{GW150914-TGR}).

The curse of dimensionality could be mitigated by multiband observations, where data from both space-based detectors, such as the Laser Interferometer Space Antenna, and ground-based observatories operating at that time are jointly analyzed for the same source~\citep{Gupta:2020lxa, Datta:2020vcj}. Unfortunately, current GW detectors lack the sensitivity required to break the correlations between various parameters. The question then becomes whether there exist interesting parameters in terms of which a multiparameter test can be cast without having to deal with the parameter degeneracies. In this context, Ref.~\cite{Mahapatra:2023uwd} proposed that a waveform model parametrizing the GW amplitude and phase in terms of the radiative multipole moments of the binary system~\citep{Kastha:2018bcr, Kastha:2019brk, Mahapatra:2023hqq, Mahapatra:2023ydi} could serve as an effective approach to enable high-precision, multiparameter tests of GR.

In the current implementation of parametrized PN tests within the LVK tests of GR, PCA can be of help since it can identify the best-measured linear combinations of the original eight PN parameters, bringing down the dimensionality of the non-GR parameter space. Despite having the advantage of dealing with lower-dimensional parameter space, the posteriors of the PCA parameters carry the information of the original multiparameter test. In other words, instead of choosing to test one of the eight PN parameters at a time, PCA allows one to optimize the multiparameter test of GR by identifying the new set of parameters that account for most of the information and have the least correlation between them.

\subsection{Review of PCA-based tests in the literature}
\label{sec:review_PCA_lit}
Reference~\cite{AP12} proposed the idea of PCA to address the problem of degeneracies caused when one looks for simultaneous departures from GR predictions in several of the PN coefficients~\cite{AIQS06a}.  Using 3.5PN accurate nonspinning phasing in the frequency domain~\cite{BDI95, BDIWW95,BDEI04,AISS05}, the authors argued that the linear combination of the original PN coefficients can be measured with significantly better precision [${\cal O}(10^{-6})$--${\cal O}(10^{-4})$] than the original PN coefficients using second-generation or third-generation ground-based detector configurations. Recently, Ref.~\cite{ShoomEtalSVD} carried out such a test in the case of GW170817~\cite{GW170817}, for which the inspiral-based PN waveform model {\tt TaylorF2}~\cite{BIOPS09}  was accurate enough, and derived the bounds on the PCA parameters. 

Reference~\cite{PCA1} revisited the problem in terms of PN {\it fractional deformation} parameters (as opposed to directly using the PN phasing coefficients as observables~\cite{AP12,ShoomEtalSVD}) that are regularly employed for testing GR using the Test Infrastructure for General Relativity ({\tt TIGER})~\cite{TIGER,Meidam:2017dgf} and Flexible Theory-Independent ({\tt FTI})~\cite{FTI} frameworks. As mentioned earlier, these are fractional additive deformations introduced at every PN order which, by definition, are zero in GR. First, they performed a multiparameter test of GR varying six out of the eight PN deformation parameters along with all the standard GR parameters in a binary black hole (BBH) signal. Then, they obtained the covariance matrix corresponding to the six-dimensional posteriors of the deformation parameters by marginalizing over the GR parameters, in the {\tt TIGER} framework. The covariance matrix was then diagonalized to obtain the new set of parameters that are linear combinations of the original deformation parameters. These are referred to as PCA parameters. The precision with which the PCA parameters are inferred was studied for a variety of simulated signals. As expected, the PCA-based method yields significantly tighter constraints on deviations from GR than the original multiparameter approach. 

When fiducial non-GR signals are injected, the PCA-based method is capable of excluding GR with high statistical confidence, provided the signal-to-noise ratio (SNR) is sufficiently large: For a non-GR signal with an SNR of 54, GR was ruled out at a confidence level of $\sim 14\,\sigma$, while an SNR of 40 yielded a $11\,\sigma$ exclusion. In contrast, a signal with an SNR of 27 led to a GR exclusion at only $\sim 4\,\sigma$. This demonstrates the method’s sensitivity to both the SNR and the injected deviation. 

Additionally, Ref.~\cite{PCA1} also carried out this test on a selected subset of GW events in the first GW transient catalog~\cite{GWTC1} and obtained the corresponding bounds on the PCA parameters.

\subsection{This work}
In this paper, we extend the work of Ref.~\cite{PCA1} by performing a comprehensive study that includes both the {\tt TIGER} and {\tt FTI} frameworks, a wider set of injections covering precession and nonquadrupole modes, and a broader range of GR and non-GR signals.
This is an important step before deploying the PCA-based test of GR (or ``{\tt PCA-TGR}'') on the real signals as they will exhibit several of these physical effects. Precisely quantifying the test’s response to a diverse set of injections is crucial for identifying potential systematic biases and for accurately interpreting results from real events---the primary objective of the present study.

More specifically, in this paper, we address the following questions:
\begin{enumerate}
     \item How does {\tt PCA-TGR} respond to different types of GR injections, which cover different physical effects present in the waveform, and how confidently does it recover GR?
    \item How well can {\tt PCA-TGR} detect a GR violation, and how does this ability depend on SNR, with analyses carried out independently using both the {\tt TIGER} and {\tt FTI} frameworks and considering a broader range of deviation scenarios, beyond the case where all PN coefficients are deformed by the same magnitude?
    \item How does  {\tt PCA-TGR} respond to unmodeled physics in the waveform, such as eccentricity? Which of the PCA parameters capture this ``false" GR violation?
    \item What do the {\tt PCA-TGR} results tell us about possible GR violation when applied to selected GWTC-3~\cite{GWTC3,GWTC3-TGR} events?
\end{enumerate}

This paper is organized as follows. Section~\ref{sec:analysis_framework} details the analysis framework. Sections \ref{sec:GRinj} and \ref{sec:non-GRinj} report the results from GR and non-GR injections, respectively. Section~\ref{sec:ecc-inj} discusses how {\tt PCA-TGR} responds to eccentric numerical relativity (NR) injections, and Sec.~\ref{sec:results_o3} discusses the results from selected events from the third observing run of LVK. Section~\ref{sec:conc} summarizes the conclusions and caveats of the method.
\section{Analysis Framework}\label{sec:analysis_framework}
In GR, the early inspiral phase of a gravitational-wave signal from a compact binary coalescence is well approximated by the PN formalism~\cite{BDI95,Bliving}. In this formalism, the amplitude and phase of gravitational waves are expressed as a series in powers of the characteristic orbital velocity $v$ of the system. The frequency domain GW phase of the inspiral part from a quasicircular compact binary, computed using the stationary phase approximation \cite{SathyaDhurandhar91, DIS00}, for the leading quadrupolar harmonic \cite{BIOPS2009, ABFO08, MKAF2016}, takes the following form in GR
\begin{align}\label{eq:PN_phase}
\Phi_{\rm insp}(f)&=2\pi f\,t_c-\phi_c-\frac{\pi}{4}+\frac{3}{128\,\eta\,v^5} 
\nonumber \\
& \times \quad\left[ \, \sum_{k=0}^{k=7} \left( \phi_k \, + \phi_{kl} \, \ln v \right) v^k \, + \mathcal{O}(v^{8}) \right] \,,
\end{align}
where $t_c$ and $\phi_c$ are the time and phase of coalescence, and $\eta$ is the symmetric mass ratio defined as the ratio of the reduced mass to the total mass of the binary. The characteristic orbital velocity parameter, in geometrical units, is given by $v  = (\pi M f)^{1/3}$, where $f$ is the GW frequency and $M$ is the (redshifted) total mass of the binary.~\footnote{We note that in the LVK papers, $(1+z)M$  is used to denote the redshifted total mass, where $z$ is the cosmological redshift and $M$ is the source-frame total mass. Here, however, we denote $M$ as the redshifted (detector-frame) total mass for brevity.}
Terms with a \textit{k}th power of $v$ are identified as the $k/2$ PN order corrections to the GW phase. The PN coefficients, denoted by $\phi_k$ (nonlogarithmic) and $\phi_{k\ell}$ (logarithmic), are unique functions of the compact binary's component masses and spins, capturing several physical effects and nonlinearities in GR. In GR, up to 3.5PN order, the logarithmic coefficients $\phi_{kl}$ are nonzero only for $k = 5, 6$.

\subsection{Multiparameter test of general relativity}
\label{sec:multiparameter_test}
In alternative theories of gravity, one or more of the PN coefficients may show deviation from GR~\cite{YYP2016,Endlich:2017tqa,Sennett:2019bpc,Bernard:2022noq}. They may have different dependencies on the component masses and spins than GR, and/or they may depend on additional parameters that characterize the alternative theory (see, for example, Ref.~\cite{Bernard:2022noq} for scalar-tensor gravity theories).

The parametrized inspiral tests search for possible deviations from the unique structure of the PN coefficients predicted by GR. This is done by introducing phenomenological, dimensionless fractional deviation parameters into the PN coefficients~\cite{MAIS10,Li:2011cg,TIGER,GW150914-TGR}:
\begin{equation}\label{eq:PN_deformation}
    \phi_{b} \longrightarrow \phi_{b}^{\rm GR} (1+\delta \hat{\phi}_{b})\,, 
\end{equation}
where the subscript ``$b$''  collectively represents both logarithmic and nonlogarithmic coefficients of Eq.~(\ref{eq:PN_phase}). This is a theory-agnostic approach wherein the parametrized fractional deviations $\delta\hat{\phi}_b$ are identically zero in GR, whereas in alternative theories of gravity, one or more of these parameters could differ from zero.
Therefore, a direct measurement of $\delta\hat{\phi}_b$ using GW data allows us to constrain any deviation from GR. Such tests are referred to as ``null'' tests of GR, with the null hypothesis being GR.

Currently, among LVK analyses, there are two different frameworks that can perform this type of GR null test. The first one is {\tt TIGER}~\cite{TIGER,TIGER-2014-Michalis}, and the second is {\tt FTI}~\cite{FTI}. The two frameworks differ in the choice of baseline GR waveform models and in how the inspiral test is implemented.  {\tt TIGER}~\cite{Roy:2025gzv} uses the {\tt IMRPhenomX} waveform family~\cite{Pratten:2020ceb,Garcia-Quiros:2020qpx,Colleoni:2024knd}, which accounts for precession, as the baseline GR waveform model. The {\tt FTI}~\cite{Sanger:2024axs} relies on the {\tt SEOBNRv4\_ROM} waveform family~\cite{Bohe:2016gbl,Cotesta:2020qhw,Cotesta:2018fcv} and is currently limited to aligned-spin systems. In the {\tt TIGER} framework, the fractional PN deviation parameters are added directly to the PN coefficients in the frequency-domain inspiral phase of the {\tt IMRPhenomX} waveform family, which ends at the minimum energy circular orbit (MECO) frequency. In the {\tt FTI} framework, the fractional PN deviation parameters are added to the full frequency-domain inspiral-merger-ringdown phase of the {\tt SEOBNRv4\_ROM} waveform family and are smoothly tapered off at the frequency corresponding to the peak of the (2,2) mode. Note that parametric fractional deviations are only applied to the nonspinning portion of the inspiral phase in both frameworks.~\footnote{This choice is motivated by the convenience of mapping the results to nonlinear effects in general relativity, such as tails, tails of tails, and so on. Since we are performing null tests, whether the deformation is applied to the total phasing or only to the nonspinning part will have negligible impact on the results. However, this distinction should be kept in mind when interpreting the outcomes of the tests.}

In the current implementation of this null test by the LVK Collaboration, only one fractional deviation parameter, $\delta\hat{\phi}_b$, 
is varied at a time while keeping the others fixed at their GR values to obtain the bounds on all the fractional deviation parameters. We refer to this approach as \emph{single-parameter tests}, which leads to a set of null tests of GR, each corresponding to a PN coefficient, although the tests are not necessarily all independent. 
While single-parameter tests can \emph{detect} deviations from GR~\cite{LiEtal2012,Sampson:2013lpa}, they\emph{cannot be uniquely attributed} to a real deviation at the corresponding PN order due to degeneracies in the waveform. In other words, it is possible that a deviation at particular PN order in a non-GR theory can be suboptimally captured by a test which looks for departures at a different PN order. Additionally, single-parameter tests ignore potential modifications of multiple PN coefficients, which can lead to an underestimation of errors and biases~\cite{Lyu:2022gdr}. As a consequence, despite their ability to detect GR deviations, the single-parameter tests provide a somewhat limited insight into the nature of alternative theories of gravity. 

In light of the limitations of single-parameter tests, we advocate for a more robust multiparameter approach to test GR, wherein multiple fractional deviation parameters are varied simultaneously~\cite{AIQS06a,LiEtal2012,AP12,TIGER-2014-Michalis}. More specifically, we consider simultaneous variation of all the fractional deformation parameters: $\{ \vec{\delta\phi}_k, \vec{\delta\phi}_{k\ell}\}$. Taking into account the inspiral phase in GR up to 3.5PN order with nonzero PN coefficients, we have a set of eight fractional deformation parameters\footnote{We do not consider fractional deviations to the nonlogarithmic term at 2.5PN order, $\delta\hat{\phi}_5$, as it is independent of frequency and can therefore be absorbed by redefining the phase at coalescence, $\phi_c$.}:
\begin{align}
    \vec{\delta\phi}_k &= \{ \delta\hat{\phi}_0, \delta\hat{\phi}_2, \delta\hat{\phi}_3, \delta\hat{\phi}_4, \delta\hat{\phi}_6, \delta\hat{\phi}_7 \}\,,\label{eq:frac_PN_deformation_1}\\
    \vec{\delta\phi}_{k\ell} &= \{ \delta\hat{\phi}_{5\ell}, \delta\hat{\phi}_{6\ell} \}\,.\label{eq:frac_PN_deformation_2}
\end{align}
The major obstacle of performing this class of tests is the uninformative posteriors due to high correlations between the parameters. We employ principal component analysis to remedy this shortcoming and identify the best measured linear combinations of the deformation parameters~\cite{AP12,Saleem:2021iwi,Datta:2020vcj,ShoomEtalSVD}. The details of the method are described below.

\subsection{Bayesian formulation}
The baseline GR waveform model for a BBH is a function of several binary parameters (denoted as $\vec{\theta}_{\rm GR}$), including component masses, spins, luminosity distance, and angular parameters describing the orientation of the binary orbit. For nonprecessing and precessing BBHs on quasicircular orbits, $\vec{\theta}_{\rm GR}$ contains 11 and 15 parameters,~\footnote{The observed GW signal emitted by quasi-circular, precessing BBHs within GR depends on 15 parameters: detector-frame masses \( m_1 \) (primary black hole) and \( m_2 \) (secondary black hole); spin magnitudes \( \chi_1 \) and \( \chi_2 \); spin-orbit tilt angles \( \theta_1 \) and \( \theta_2 \) (angles between the spin vectors and the orbital angular momentum); spin azimuthal angle difference \( \phi_{12} \) (the difference between the azimuthal angles of the two spin vectors); \( \phi_{JL} \) (the angle between the orbital and total angular momentum); inclination angle \( \theta_{JN} \) (between the line of sight and total angular momentum); luminosity distance \( D_L \); sky location (right ascension $\alpha$ and declination $\delta$); polarization angle \( \psi \); time of coalescence, \( t_c \); and phase at coalescence, \( \phi_c \). For aligned-spin binaries, only two spin parameters (\( \chi_1, \chi_2 \)) are relevant, reducing the total number of parameters to 11.} respectively. In addition to $\vec{\theta}_{\rm GR}$, the parametrized waveform models under consideration are also functions of eight fractional PN deviation parameters: $\vec{\theta}_{\rm D}=\{ \vec{\delta\phi}_k, \vec{\delta\phi}_{k\ell}\}$. In general, the parametrized waveform models depend on $m$ GR parameters and $n$ fractional deviation parameters. 

In a multiparameter test of GR, this implies running a Bayesian parameter estimation algorithm to infer $m$ GR parameters $\vec{\theta}_{\mathrm{GR}}$ and $n$ non-GR parameters $\vec{\theta}_{\mathrm{D}}$. Given a hypothesis $\mathcal{H}$ that the data $d_j$ of event $j$ are a sum of Gaussian noise and a GW signal described by a waveform model that is a function of $\vec{\theta}_{\mathrm{GR}}$ and $\vec{\theta}_{\mathrm{D}}$ parameters, the posterior probability distribution of parameters $\vec{\theta}_{\mathrm{GR}}$ and $\vec{\theta}_{\mathrm{D}}$ can be written using Bayes' theorem as
\begin{equation}
    \label{eq:Bayes_theorem}
    p(\{\vec{\theta}_{\mathrm{GR}}, \vec{\theta}_{\mathrm{D}}\} | d_j, \mathcal{H}) = 
\frac{\mathcal{L}(d_j | \{\vec{\theta}_{\mathrm{GR}}, \vec{\theta}_{\mathrm{D}}\}, \mathcal{H}) \, \pi(\{\vec{\theta}_{\mathrm{GR}}, \vec{\theta}_{\mathrm{D}}\} | \mathcal{H})}{\mathcal{Z}(d_j | \mathcal{H})}\,.
\end{equation}
Here, $\pi(\{\vec{\theta}_{\mathrm{GR}}, \vec{\theta}_{\mathrm{D}}\} | \mathcal{H})$ is the \emph{prior probability} distribution of  $\vec{\theta}_{\mathrm{GR}}$ and $\vec{\theta}_{\mathrm{D}}$, $\mathcal{L}(d_j | \{\vec{\theta}_{\mathrm{GR}}, \vec{\theta}_{\mathrm{D}}\}, \mathcal{H})$ is the \emph{likelihood function} and $\mathcal{Z}(d_j | \mathcal{H})$ is the \emph{Bayesian evidence} in favor of the hypothesis $\mathcal{H}$. To generate discrete samples for the probability distribution function $p(\{\vec{\theta}_{\mathrm{GR}}, \vec{\theta}_{\mathrm{D}}\} | d_j, \mathcal{H})$, we used the Bayesian parameter inference package {\tt Bilby TGR}~\citep{bilby_tgr} based on {\tt Bilby}~\citep{bilby_paper,bilby_pipe_paper}, with the {\tt Dynesty}~\citep{dynesty} sampler (which uses the nested sampling algorithm~\cite{Skilling}). We adopt the prior probability distribution $\pi(\{\vec{\theta}_{\mathrm{GR}}, \vec{\theta}_{\mathrm{D}}\} | \mathcal{H})$ 
following the standard setup used in the publicly available LVK data~\cite{GWTC3,lvc:dataGWTC3,GWTC3-TGR}. This means for the GR sector the detector-frame component masses and spin magnitudes of individual black holes have uniform priors, while spin orientations are distributed isotropically over a sphere. The luminosity distance $D_L$ is assigned a prior so that sources are uniformly distributed in the comoving volume and source-frame time. The priors for all other angular parameters are chosen to be isotropic over a sphere. In general, the chosen priors of the standard GR parameters are uninformative and sufficiently broad to encompass the regions of the parameter space where the posterior distributions are supported while also balancing computational efficiency and remaining within the calibration range of the waveform models. For the  non-GR parameters, we assume a uniform symmetric prior between $-20$ and $20$, centered at zero.

As our focus is on the fractional deformation parameters, we marginalize Eq.~\eqref{eq:Bayes_theorem} over the $m$ GR parameters to obtain $n$-dimensional joint posterior distribution of $\vec{\theta}_{\mathrm{D}}$,
\begin{equation}
    \label{eq:marginalize_GR}
    p(\vec{\theta}_{\mathrm{D}} | d_j, \mathcal{H}) = 
\int p(\{\vec{\theta}_{\mathrm{GR}}, \vec{\theta}_{\mathrm{D}}\} | d_j, \mathcal{H}) \, d\vec{\theta}_{\mathrm{GR}}.
\end{equation}
Moreover, we only vary six deviation parameters from 1.5PN to 3.5PN, fixing the 0PN and 1PN deviation parameters to their GR value of zero, leading to $\vec{\theta}_{\rm D} = \{ \delta\hat{\phi}_3, \delta\hat{\phi}_4, \delta\hat{\phi}_{5\ell}, \delta\hat{\phi}_6, \delta\hat{\phi}_{6\ell}, \delta\hat{\phi}_7 \}$. The reason for this is explained at the end of Sec.~\ref{sec:PCA}. We do not consider $-1$PN and $0.5$PN deviations, which we set to zero.

\subsection{Principal component analysis}\label{sec:PCA}
As noted earlier, the fractional deformation parameters $\vec{\theta}_{\mathrm{D}}$ are generally correlated. Consequently, a multiparameter test using the fractional deformation parameters described in the previous section can lead to degenerate posteriors, resulting in a nondiagonal covariance matrix in their joint distribution. To address this, we aim to construct a new set of orthogonal parameters by rotating the axes of the original fractional deformation parameters such that the resulting covariance matrix becomes diagonal. This is accomplished by using the principal component analysis, following the method outlined in Ref.~\cite{PCA1}, with the key steps summarized below.
\begin{enumerate}[label=(\roman*)]
    \item  We compute the variance-covariance matrix for fractional deviation parameters from the $\vec{\theta}_{\mathrm{D}}$ posteriors obtained in Eq.~\eqref{eq:marginalize_GR},
    \begin{equation}
        C_{ab} = \left\langle \left( \delta\hat{\phi}_a - \langle \delta\hat{\phi}_a \rangle \right) \left( \delta\hat{\phi}_b - \langle \delta\hat{\phi}_b \rangle \right)\right\rangle\,,
    \end{equation}
    where the angular brackets denote averaging over the marginalized posterior distribution.
    \item 
    We diagonalize the covariance matrix $C$ as
    \begin{equation}
        C = USU^T,
    \end{equation}
    where $S$ is the diagonal matrix containing the eigenvalues, and $U$ is the unitary matrix whose columns are the corresponding eigenvectors representing the transformation of the basis vectors. Here, we use the {\tt numpy.linalg.svd} function in Python~\cite{Harris:2020xlr} to diagonalize the covariance matrix—or more specifically, to obtain the $U$ and $S$ matrices.
    \item 
    In this context, $S$ represents a diagonal variance-covariance matrix corresponding to the new deformation parameters, which are obtained by applying the transformation matrix $U$ (with elements $\alpha_{ib}$) on the original deformation parameters,
    \begin{equation}\label{eq:pca_comp}
        \delta\hat{\phi}^{(i)}_{\rm PCA} = \sum_b \alpha_{ib} \delta\hat{\phi}_b\,, 
    \end{equation}
    which we refer to as the PCA parameters.
    \item PCA has an inherent sign ambiguity because the direction of eigenvectors is not uniquely defined; flipping the sign of an eigenvector does not change the variance it explains. In other words, both $\vec{u}_{i}$ and $-\vec{u}_{i}$ are valid eigenvectors of the covariance matrix $C$, yielding the same diagonalized form $S$. This ambiguity carries over to the principal components $\delta\hat{\phi}^{(i)}_{\rm PCA}$, potentially causing inconsistencies in interpretation across analyses of different GW events. To resolve this, we adopt a sign convention in which each eigenvector, and the corresponding principal component, is adjusted based on the sign of the eigenvector’s largest (in magnitude) element—ensuring it is always positive. This choice does not affect the statistical properties of the PCA parameters but guarantees consistency in the orientation of the new basis. As a result, the PCA parameters remain well defined and on a comparable footing across different GW events in our analysis.

    \item 
    We perform numerical consistency checks to validate the diagonalization procedure.   
    First, we assess the \textit{orthonormality} of the transformation matrix $U$ by evaluating the maximum absolute value of the off-diagonal elements of $UU^\mathrm{T} - \mathbb{I}$:
    \begin{equation}
        \max_{a \ne b} \left| (UU^\mathrm{T} - \mathbb{I})_{ab} \right|,
    \end{equation}
    where $(\mathbb{I})_{ab}$ denotes the element in the $a$th row and $b$th column of the identity matrix. 
    This should vanish for a perfectly orthonormal matrix. Second, we reconstruct the covariance matrix using $C = USU^\mathrm{T}$ and compute the Frobenius norm of the difference with the original covariance matrix:
    \begin{equation}
        \left\| USU^\mathrm{T} - C \right\|_F,
    \end{equation}
    to quantify the total reconstruction error introduced by the numerical diagonalization.

    In all cases applied here, we find that the maximum off-diagonal element of $|UU^\mathrm{T}|$ is $\mathcal{O}(10^{-16})$, and the Frobenius norm of $USU^\mathrm{T} - C$ is $\mathcal{O}(10^{-13})$, confirming the numerical stability and internal consistency of the PCA-based diagonalization.

    \item 
    The primary advantage of the PCA formalism is dimensionality reduction. In our case, this is guided by examining the eigenvalues, i.e., the diagonal elements of the matrix $S$. Each eigenvalue corresponds to the variance (squared uncertainty) of the posterior along the direction defined by the associated eigenvector in $U$. A hierarchy naturally emerges: eigenvectors associated with smaller eigenvalues correspond to directions with tighter constraints and, hence, carry more information. Conversely, directions with large eigenvalues (i.e., large error bars) are poorly constrained and largely uninformative—often yielding posteriors that resemble the priors. Therefore, such directions can be safely excluded from further analysis, enabling a truncated, lower-dimensional representation of the parameter space depending on the acceptable tolerance in reproducing the data with the lower dimensional PCA parameters.
    
    \item 
    The information content encoded in the posterior distribution can be assessed using the inverse of the covariance matrix, also known as the information matrix. The overall information can then be quantified by computing the determinant of the inverse of $S$. Previous studies \cite{PCA1, PCA2} have shown that retaining only the two most tightly constrained principal components often captures more than 99\% of the total information. In this work, rather than relying solely on eigenvalue hierarchy, we also use the Jensen–Shannon (JS) divergence~\cite{JSdiv} to determine the number of PCA parameters to retain. Specifically, we retain only those PCA components for which the JS divergence between the posterior and prior distributions exceeds 0.1 bits, ensuring that the selected components carry meaningful information gained from the data.

\end{enumerate}

Equation~\eqref{eq:pca_comp} defines a new set of dimensionless deformation parameters, each of which is a linear combination of the original PN fractional deviation parameters. These transformed parameters effectively encapsulate the joint information content of a multiparameter test.  
The approach outlined above is data-driven, meaning that the new parameters are derived directly from the posterior distribution, making the resulting parametrization inherently unique to each event.
Because the posterior is unique to each event, this motivates the use of a data-driven parametrization, as it allows the method to adapt to the information content of each event while avoiding assumptions about the underlying parameter values.
An alternative strategy involves estimating the covariance matrix using the Fisher information matrix formalism~\cite{Rao45,Cramer46,CF94,PW95}. However, this requires prior knowledge of the true parameter values—information that is generally inaccessible, especially in scenarios where deviations from GR might be present. Furthermore, preliminary investigations have shown inconsistencies between Fisher matrix predictions and the outcomes of the data-driven approach, likely due to the approximations and limitations inherent in the Fisher formalism for the SNRs of typical LVK events, such as bimodality and other non-Gaussianities in the posteriors.

In the current formulation of the PCA-based multiparameter test, we exclude 0PN and 1PN fractional deformation parameters and instead vary coefficients from 1.5PN to 3.5PN. This is due to the limited sensitivity of current GW detectors, which cannot effectively disentangle these lower-order fractional deformations from intrinsic binary parameters like the chirp mass and symmetric mass ratio. Furthermore, the lower PN deviation parameters are already tightly constrained by binary pulsar observations~\cite{Yunes:2010qb,Kramer:2021jcw}. 
For this reason, the current analysis is limited to a six-dimensional multiparameter test, with deformation parameters introduced at orders ranging from 1.5PN to 3.5PN. However, with future improvements in detector sensitivity or the observation of high-SNR events with longer inspirals, the PCA framework can be extended to perform a full eight-dimensional test, enabling a more complete exploration of PN deformations.

\subsection{Combining information from multiple events}
\label{subsec:comb_events}
The PCA parameters, obtained by the procedure outlined above, will be different for different events. Therefore, joint constraints on the PCA parameters have to be clearly defined. There are two commonly used approaches for combining results from single-parameter tests of GR across multiple GW events. 
The first approach, which we adopt in this work, involves multiplying the individual likelihoods of the deformation parameters obtained for each event~\cite{DelPozzo:2011pg,Li:2011cg}. This method assumes that the value of each deformation parameter is identical across all events, an admittedly strong and potentially unrealistic assumption given the variation in source properties, but one that is valid if GR holds true.  
An alternative approach involves hierarchical inference~\cite{Zimmerman:2019wzo,Isi:2019asy}, which models event-to-event variation through a population distribution, typically modeled as Gaussian.

In the context of PCA-based multiparameter tests, combining results across events presents additional challenges. Unlike single-parameter analyses, PCA parameters are event-specific due to differences in the transformation matrix that defines them. As a result, the same PCA index corresponds to different linear combinations of the original fractional PN deviation parameters across events, making the post-PCA combination technically inconsistent.
We therefore combine events at the level of the original multiparameter posteriors—using the shared-parameter likelihood multiplication method adopted in this work—and then perform PCA on the combined posterior. While a hierarchical approach could, in principle, be extended to this setting, modeling the joint distribution of six correlated parameters would require estimating 27 hyperparameters under the Gaussian model assumption, a task not yet explored in high-dimensional GW analyses. We leave such developments to future work.

To combine data from a set of, say, $N$ events, we begin by computing the six-dimensional marginalized likelihood for the deformation parameters, $\vec{\theta}_{\rm D}$, for each event, following Eqs.~(\ref{eq:Bayes_theorem}) and~(\ref{eq:marginalize_GR}). The combined likelihood is then constructed as the product of the $N$ multidimensional likelihoods:
\begin{equation}
\mathcal{L}^{\mathrm{combined}}\left( \{d_j\}\mid\vec{\theta}_{\mathrm{D}}, \mathcal{H} \right) = \prod_{j=1}^{N} \mathcal{L}\left(d_j\mid\vec{\theta}_{\mathrm{D}}, \mathcal{H}\right),
\label{eq:combined-likelihood-non-pca}
\end{equation}
where each \(\mathcal{L}(d_j\mid\vec{\theta}_{\mathrm{D}}, \mathcal{H})\) is computed by dividing the marginalized posterior distribution in Eq.~(\ref{eq:marginalize_GR})  
by the prior on $\vec{\theta}_{\mathrm{D}}$, \(\pi(\vec{\theta}_{\mathrm{D}}\mid \mathcal{H})\). Given that uniform priors were used for all six parameters in the original analyses, the likelihoods are directly proportional to the posteriors, and because the posteriors of each event are provided as discrete samples, the likelihoods are constructed by sampling from the six-dimensional kernel density estimator fits using the {\tt sklearn.mixture.GaussianMixture} function~\cite{scikit-learn}. 
The combined posterior on $\vec{\theta}_{\rm D}$ is then obtained by multiplying the combined likelihood, as given in Eq.~(\ref{eq:combined-likelihood-non-pca}), with the same uniform prior on $\vec{\theta}_{\rm D}$ used in the original single-event analysis:
\begin{equation}
p\left( \vec{\theta}_{\mathrm{D}}\mid\{d_j\}, \mathcal{H} \right) \propto \pi(\vec{\theta}_{\mathrm{D}} \mid \mathcal{H}) \, \mathcal{L}^{\mathrm{combined}}\left( \{d_j\} \mid \vec{\theta}_{\mathrm{D}}, \mathcal{H} \right)\,.
\label{eq:combined-posterior-non-pca}
\end{equation}
Furthermore, to generate discrete samples for the probability distribution function $p\left( \vec{\theta}_{\mathrm{D}}\mid\{d_j\}, \mathcal{H} \right)$, we again used the Bayesian parameter inference library {\tt bilby}~\citep{bilby_paper} with the {\tt dynesty}~\citep{dynesty} sampler.
Finally, the PCA formalism, described in Sec.~\ref{sec:PCA}, is applied to the discrete samples of the combined six-dimensional posterior probability distribution $p\left( \vec{\theta}_{\mathrm{D}}\mid\{d_j\}, \mathcal{H} \right)$ to obtain the joint constraints on the PCA parameters.

After describing the analysis framework of the {\tt PCA-TGR}, we validate its efficacy through various types of injection studies, as discussed below.

%%%%%%%%%%%%%%%%%%%%%%%%%%%%%%%%%%%%%%%%%%%%%%%%%%%%%%%%%%%%%
\begingroup
\setlength{\tabcolsep}{0.75pt} % Default value: 6pt
\renewcommand{\arraystretch}{1.25} % Default value: 1
\begin{table*}[t]
\begin{center}
\begin{tabular}{|c|c|c|c|c|c|c|c|c|c|}
%\begin{tabular}{|p{0.05cm}|p{2cm}|p{0.5cm}|p{0.5cm}| p{0.5cm}|p{0.5cm}|p{0.5cm}|p{0.5cm}|p{0.5cm}|p{0.5cm}| p{0.5cm}|}
\hline
\multicolumn{1}{| c |}{ No.} & \multicolumn{5}{c|}{Properties of GR injections} & \multicolumn{2}{c|}{Properties of $\delta \hat{\phi}_{\rm PCA}^{(1)}$} & \multicolumn{2}{c|}{Properties of $\delta \hat{\phi}_{\rm PCA}^{(2)}$}\\[0.2cm]
\hline
 & $M$  & $q$ & ($\chi_1,\chi_2$) & $D_L$  & SNR & Median and  & $\mathcal{Q}_{\rm GR}$ & Median and & $\mathcal{Q}_{\rm GR}$ \\[0.2cm]
 & ($M_{\odot}$) & & & (Mpc) & & 90\% errors & ($z_{\rm GR}$) & 90\% errors & ($z_{\rm GR}$)\\[0.2cm]
 \hline
 \hline
 1. & 25 & 1.5 & (0.2,0.1) & 1200 & $ 20$ & $-0.02_{-0.21}^{+0.17}$ & 0.57 (0.16) & $-0.09_{-1.0}^{+0.8}$ & 0.57 (0.15) \\[0.2cm]
    &    &     &            & 650 & $ 40$ & $-0.0_{-0.1}^{+0.1}$ & 0.52 (0.05) & $0.0_{-0.43}^{+0.39}$ & 0.5 (0.01) \\
\hline
2. & 25 & 1.5 & (0.8,0.7) & 1400 & $ 20$ & $-0.07_{-0.26}^{+0.11}$ & 0.76 (0.66) & $-0.18_{-1.37}^{+0.81}$ & 0.63 (0.26) \\[0.2cm]
   &    &     &            & 700 & $ 40$ & $-0.04_{-0.13}^{+0.06}$& 0.77 (0.73) & $-0.09_{-0.64}^{+0.44}$ & 0.63 (0.28)  \\
\hline
3. & 25 & 4 & (0.2,0.1) & 900 & $ 20$ & $-0.01_{-0.2}^{+0.21}$& 0.53 (0.06) & $0.12_{-1.14}^{+1.3}$ & 0.42 (0.16)\\[0.2cm]
   &    &    &           & 500 & $ 40$ & $-0.0_{-0.12}^{+0.12}$ & 0.5 (0.0) & $0.04_{-0.76}^{+1.06}$ & 0.46 (0.07) \\
\hline
4. & 25 & 4 & (0.8,0.7) & 1000 & $ 20$ & $-0.04_{-0.22}^{+0.13}$ & 0.66 (0.34) & $-0.21_{-1.42}^{+0.94}$ & 0.62 (0.24) \\[0.2cm]
   &    &    &          & 550   & $ 40$ & $-0.03_{-0.1}^{+0.04}$ & 0.75 (0.62)& $-0.02_{-0.65}^{+0.51}$ & 0.53 (0.06)\\
\hline
\hline
5. & 60 & 1.5 & (0.2,0.1) & 2400 & $ 20$ & $-0.04_{-0.29}^{+0.22}$& 0.6 (0.22) & $-0.18_{-2.25}^{+1.57}$ & 0.56 (0.15) \\[0.2cm]
   &    &     &           & 1200 & $ 40$ & $-0.0_{-0.15}^{+0.16}$ & 0.5 (0.01) & $-0.04_{-1.04}^{+1.18}$ & 0.52 (0.06)\\
\hline
6. & 60 & 1.5 & (0.8,0.7) & 2800 & $ 20$ & $0.0_{-0.28}^{+0.3}$ & 0.49 (0.02) & $0.59_{-0.73}^{+1.98}$&0.23 (0.71) \\ [0.2cm]
   &    &     &           & 1400 & $ 40$ & $0.02_{-0.19}^{+0.23}$& 0.45 (0.13)&$0.21_{-0.47}^{+0.87}$ & 0.29 (0.52)\\
 \hline
7. & 60 & 4 & (0.2,0.1) & 1800 & $ 20$ & $-0.02_{-0.34}^{+0.36}$ & 0.54 (0.09) & $0.32_{-2.21}^{+2.38}$ & 0.41 (0.22) \\ [0.2cm]
   &    &   &           & 950  & $ 40$ & $0.0_{-0.17}^{+0.21}$ & 0.5 (0.0) & $0.21_{-1.34}^{+1.68}$&0.41 (0.23)  \\
 \hline
8. & 60 & 4 & (0.8,0.7) & 2100 & $ 20$ & $0.01_{-0.26}^{+0.29}$& 0.48 (0.03)& $-0.11_{-1.43}^{+1.17}$& 0.56 (0.13) \\[0.2cm]
   &    &   &           & 1100 & $ 40$ & $-0.01_{-0.17}^{+0.16}$& 0.55 (0.11) & $0.03_{-0.7}^{+0.83}$& 0.47 (0.06)\\   
\hline
\end{tabular}
\caption{Summary of results from the simulated aligned-spin BBH GR injections study within the {\tt TIGER} framework, demonstrating the efficacy of the {\tt PCA-TGR} method as discussed in Sec.~\ref{sec:align_spin_inj}. We simulate GW signals using the {\tt IMRPhenomXAS} waveform model, and perform parameter estimation with the {\tt parametrized IMRPhenomXAS} model. The specifications of the binary systems and the network SNRs (rounded to the nearest integer) of the injected signals are listed. For each injection, the table shows the median value, as well as the upper and lower limits of the 90\% credibility interval, $\mathcal{Q}_{\rm GR}$, and $z_{\rm GR}$ for the posteriors of $\delta\hat{\phi}^{(1)}_{\rm PCA}$ and $\delta \hat{\phi}^{(2)}_{\rm PCA}$. 
}
\label{tab:table_gr_inj_aligned_spin}
\end{center}
\end{table*}
\endgroup
%%%%%%%%%%%%%%%%%%%%%%%%%%%%%%%%%%%%%%%%%%%%%%%%%%%%%%%%%%%%%
%%%%%%%%%%%%%%%%%%%%%%%%%%%%%%%%%%%%%%%%%%%%%%%%%%%%%%%%%%%%%

\section{Response of PCA-TGR to GR Injections}\label{sec:GRinj}
First, we briefly summarize the key characteristics common to all injections discussed in this section and the following two sections (Secs.~\ref{sec:non-GRinj} and~\ref{sec:ecc-inj}). Unless otherwise specified, all reported injections assume the ‘zero-noise’ approximation in a three-detector LIGO-Virgo network~\cite{Acernese_2015,Aasi_2015,KAGRA:2013rdx} using the predicted noise power spectral densities at design sensitivity: {\tt aLIGOZeroDetHighPower} for the LIGO detectors~\cite{ALIGO_det} and {\tt AdvVirgo} for the Virgo detector~\cite{AVirgo_det}. This approximation is used to isolate potential systematic biases and to avoid statistical fluctuations arising from specific noise realizations. A lower cutoff frequency of 20 Hz is used in the computation of the likelihood, while the upper cutoff frequency varies across systems, primarily based on their masses. These values are chosen to ensure that no signal content is lost, at the same time avoiding unnecessarily high computational costs arising from arbitrarily large cutoffs. All other parameter estimation settings---such as signal duration, sampling frequency, and sampler configurations---generally follow the standard setup used in previous LVK publications~\cite{GWTC3,lvc:dataGWTC3,GWTC3-TGR}. It is worth noting that signal duration and sampling frequency are system dependent, chosen based on the physical properties of the injections. In contrast, the sampler settings remain largely consistent across all analyses.

To quantify if the posterior distributions of deformation parameters include or exclude the GR value, for each injection we compute the GR quantiles $\mathcal{Q}_{\rm GR}=P(\delta \hat{\phi}_{\rm PCA}^{(i)}<0)$ of the PCA posterior distributions. A value of $\mathcal{Q}_{\rm GR}$ significantly different from 0.5 indicates that the GR hypothesis falls in the tail of the distribution. We also compute the GR $z$ score ($z_{\rm GR}$), defined as the number of standard deviations by which the GR value (zero) deviates from the statistical median of the posterior.

\subsection{TIGER framework}
\subsubsection{Aligned-spin systems}
\label{sec:align_spin_inj}

%%%%%%%%%%%%%%%%%%%%%%%%%%%%%%%%%%%%%%%%%%%%%%%%%%%%%%%%%%%%%%%%%%%%
\begin{figure*}[hbt!]
    \centering
\includegraphics[width=0.4955\textwidth]{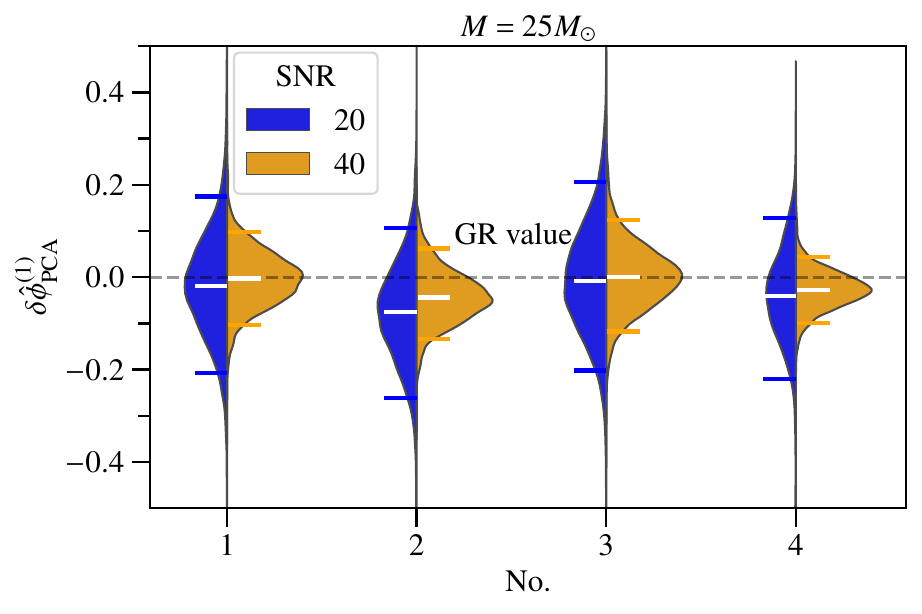}
\includegraphics[width=0.4955\textwidth]{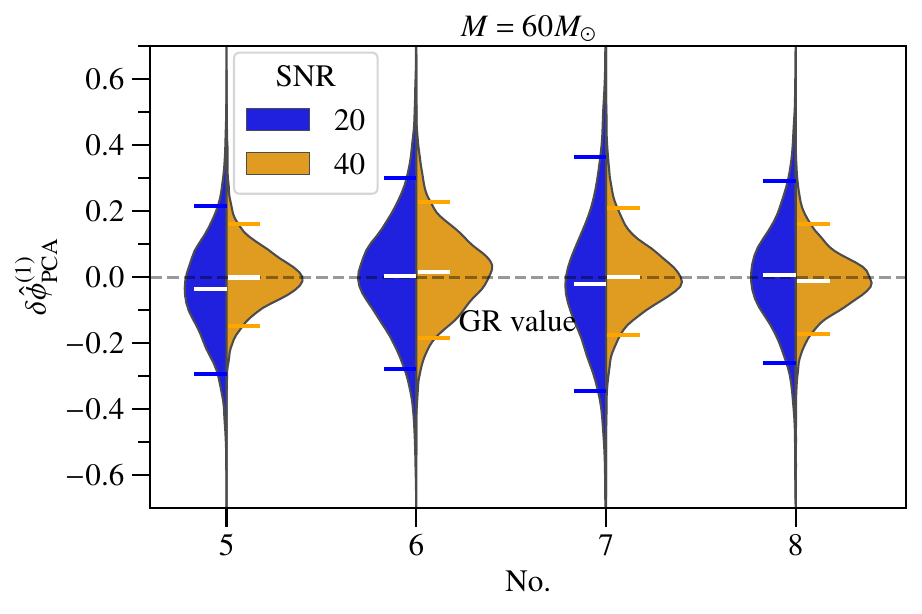}\\
\includegraphics[width=0.4955\textwidth]{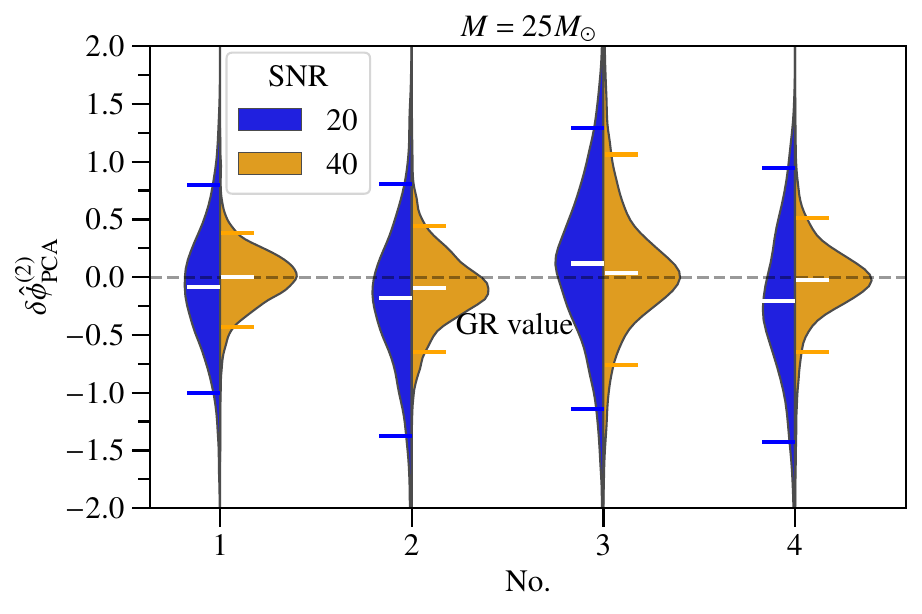}
\includegraphics[width=0.4955\textwidth]{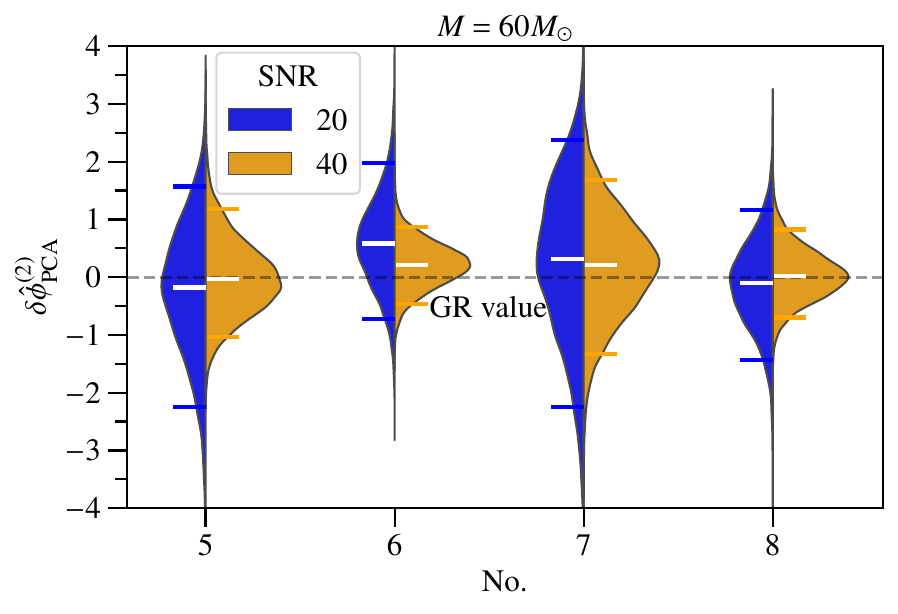}
    \caption{Posteriors of $\delta\hat{\phi}^{(1)}_{\rm PCA}$ (upper panel) and  $\delta\hat{\phi}^{(2)}_{\rm PCA}$ (lower panel) for each simulated injection—listed in Table~\ref{tab:table_gr_inj_aligned_spin} and discussed in Sec.~\ref{sec:align_spin_inj}—with the left and right panels displaying results from BBHs with detector-frame total masses of $25 M_{\odot}$ and $60 M_{\odot}$, respectively. In each violin plot, the colored horizontal bars and the horizontal white solid line denote the 90\% credible intervals and the posterior median, respectively. We mark the GR value of zero with dashed gray lines. 
    }
    \label{fig:gr_inj_aligned_spin}
\end{figure*}
%%%%%%%%%%%%%%%%%%%%%%%%%%%%%%%%%%%%%%%%%%%%%%%%%%%%%%%%%%%%%%%%
%%%%%%%%%%%%%%%%%%%%%%%%%%%%%%%%%%%%%%%%%%%%%%%%%%%%%%%%%%%%%%

Our first goal is to assess the efficacy of the {\tt PCA-TGR} in response to GR-consistent signals across different regions of the BBH parameter space (i.e., varying total masses, mass ratios, and spin magnitudes, and network SNRs). Because of the computational cost of multiparameter analyses, we begin with a simplified setup using aligned-spin quadrupolar waveform models to reduce computational overhead. In later sections, however, we incorporate more complex models that include spin precession and higher-order modes.

Results for the GR-consistent, aligned-spin injections are shown in Table~\ref{tab:table_gr_inj_aligned_spin} and Fig.~\ref{fig:gr_inj_aligned_spin}. As shown in Table~\ref{tab:table_gr_inj_aligned_spin}, we consider binaries with fixed intrinsic parameters evaluated at two network SNRs: $\sim$20 (low) and $\sim$40 (high), achieved by adjusting the luminosity distance. The extrinsic parameters---such as sky position, polarization angles, and phase at coalescence---are chosen randomly. To span a broad parameter space, we use total detector-frame masses of $25 M_{\odot}$ and $60 M_{\odot}$, with mass ratios $q = 1.5$ and 4. For spin magnitudes we choose $(\chi_1,\chi_2) = (0.2,0.1)$ and $(0.8,0.7)$ as representative low- and high-spin configurations, respectively.
These combinations yield eight distinct binary configurations, each analyzed at both low and high network SNRs, as listed in Table \ref{tab:table_gr_inj_aligned_spin}.
We use the {\tt IMRPhenomXAS} waveform model~\cite{Pratten:2020fqn} to generate injected GW signals consistent with GR and employ the {\tt parametrized IMRPhenomXAS} model~\cite{Roy:2025gzv} for parameter estimation, producing discrete samples of the posterior probability distribution given in Eq.~(\ref{eq:Bayes_theorem}) within the {\tt TIGER} framework. We then perform PCA on the marginalized six-dimensional posteriors of the fractional PN deviation parameters to construct the dominant linear combinations that are best measured, as described in Sec.~\ref{sec:PCA}.

We find that the JS divergence between the posterior and prior distributions exceeds 0.1 bits\footnote{When the JS divergence is 0 bit, the two distributions are identical, and when it is 1 bit, the two distributions are maximally different.} for the first two leading PCA parameters in all injections, while it is marginally greater than 0.1 bits for the third leading PCA parameter for some cases. 
As expected, due to the nature of PCA, the widths of the first two dominant PCA parameters, $\delta\hat{\phi}^{(1)}_{\rm PCA}$ and $\delta\hat{\phi}^{(2)}_{\rm PCA}$, are significantly smaller than those of the original fractional PN deformation parameters. 

%%%%%%%%%%%%%%%%%%%%%%%%%%%%%%%%%%%%%%%%%%%%%%%%%%%%%%%%%%%%%%%%%%%%%%%%%%%%
\begingroup
\setlength{\tabcolsep}{0.75pt} % Default value: 6pt
\renewcommand{\arraystretch}{1.25} % Default value: 1
\begin{table*}[ht]
\begin{center}
\begin{tabular}{|c|c|c|c|c|c|c|c|c|c|c|c|}
\hline
\multicolumn{1}{| c |}{ No.} & \multicolumn{7}{c|}{Properties of GR injections} & \multicolumn{2}{c|}{Properties of $\delta \hat{\phi}_{\rm PCA}^{(1)}$} & \multicolumn{2}{c|}{Properties of $\delta \hat{\phi}_{\rm PCA}^{(2)}$}\\[0.1cm]
\hline
 & $M$  & \  $q$\ \  & ($\chi_1,\chi_2$) & ($\theta_1$,$\theta_2$) & $\chi_p$ & $ D_L$  & SNR & Median and  & $\mathcal{Q}_{\rm GR}$ & Median and & $\mathcal{Q}_{\rm GR}$ \\[0.1cm]
 & ($M_{\odot}$) & & & & & (Mpc) & & 90\% errors & ($z_{\rm GR}$) & 90\% errors & ($z_{\rm GR}$)\\[0.2cm]
 \hline
 \hline
 1. & 30 & 2 & (0.3,0.2) & ($\pi/9$,$\pi/4$) & 0.10 & 1200 & 20 & $0.0^{+0.18}_{-0.18}$ & 0.49 (0.02) & $-0.08^{+0.95}_{-0.97}$ & 0.56 (0.14) \\[0.1cm]
    &    &     &    &    &    & 650 & 40 & $0.02^{+0.11}_{-0.13}$ & 0.41 (0.21) & $-0.05^{+0.59}_{-0.6}$ & 0.56 (0.15) \\
\hline
 2. & 30 & 2 & (0.8,0.7) & ($\pi/3$,$\pi/4$) & 0.69 & 1200 & 20 & $-0.04^{+0.2}_{-0.19}$ & 0.63 (0.29) & $-0.19^{+0.97}_{-0.93}$ & 0.64 (0.32) \\[0.1cm]
    &    &     &       &    &   & 650 & 40 & $-0.02^{+0.12}_{-0.11}$ & 0.61 (0.27) & $-0.09^{+0.58}_{-0.56}$ & 0.62 (0.27) \\
\hline
\hline
\end{tabular}
\caption{
Summary of results from the simulated precessing-spin BBH GR injection study conducted within the {\tt TIGER} framework, as described in Sec.~\ref{subsec:prec_gr_injs}. GW signals are simulated using the {\tt IMRPhenomXP} waveform model, and parameter estimation is performed using the {\tt parameterized IMRPhenomXP} model. The statistics of the two leading PCA parameters are reported, and their corresponding posterior distributions are shown in Fig.~\ref{fig:GR_inj_prec}.
}
\label{tab:table_gr_inj_precess_spin}
\end{center}
\end{table*}
\endgroup
%%%%%%%%%%%%%%%%%%%%%%%%%%%%%%%%%%%%%%%%%%%%%%%%%%%%%%%%%%%%%%%%%%%%%%%%%%%%
%%%%%%%%%%%%%%%%%%%%%%%%%%%%%%%%%%%%%%%%%%%%%%%%%%%%%%%%%%%%%%%%%%%%%%%%%%%%

%%%%%%%%%%%%%%%%%%%%%%%%%%%%%%%%%%%%%%%%%%%%%%%%%%%%%%%%%%%%%%%%%%%%%%%%%%%%
%%%%%%%%%%%%%%%%%%%%%%%%%%%%%%%%%%%%%%%%%%%%%%%%%%%%%%%%%%%%%%%%%%%%%%%%%%%%
\begin{figure*}[t]
 \centering
    \includegraphics[width=0.98\linewidth]{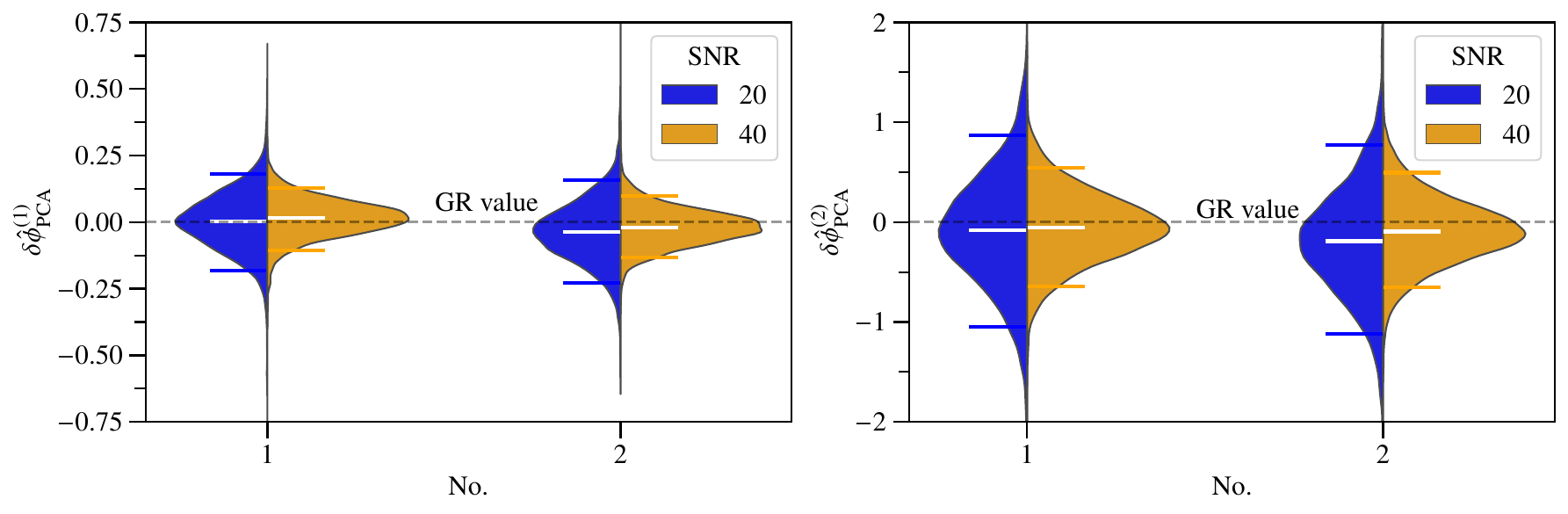}
    \caption{
    Posterior distributions of $\delta\hat{\phi}^{(1)}_{\rm PCA}$ (left panel) and  $\delta\hat{\phi}^{(2)}_{\rm PCA}$ (right panel) for each simulated precessing-spin GR injection---listed in Table~\ref{tab:table_gr_inj_precess_spin} and discussed in Sec.~\ref{subsec:prec_gr_injs}---presented as violin plots. 
    The colored and white horizontal bars have the same meaning as in Fig.~\ref{fig:gr_inj_aligned_spin}. For all the binary configurations, both the PCA parameters are consistent with the GR value (dashed gray line) of 0.}
    \label{fig:GR_inj_prec}
\end{figure*}
%%%%%%%%%%%%%%%%%%%%%%%%%%%%%%%%%%%%%%%%%%%%%%%%%%%%%%%%%%%%%%%%%%%%%%%%%%%%
%%%%%%%%%%%%%%%%%%%%%%%%%%%%%%%%%%%%%%%%%%%%%%%%%%%%%%%%%%%%%%%%%%%%%%%%%%%%

In Fig.~\ref{fig:gr_inj_aligned_spin}, we show the posterior distributions for the leading two PCA parameters presented as violin plots for each injection analysis. 
Each binary configuration is labeled by a number on the $x$-axis, corresponding to the list of binaries in Table~\ref{tab:table_gr_inj_aligned_spin}. In Table~\ref{tab:table_gr_inj_aligned_spin}, we also report the median
and 90\% credible intervals, along with $\mathcal{Q}_{\rm GR}$ and $z_{\rm GR}$ for the posteriors of the two leading PCA parameters.
First, we observe that the GR value is well recovered in all cases, as zero lies within the 90\% credible interval. This confirms that {\tt PCA-TGR} can successfully recover GR across different binary configurations. Second, as expected, the constraints on the PCA parameters improve with increasing SNR. We also find that the constraints are slightly tighter for $q=4$ compared to $q=1.5$, which can be attributed to the larger number of signal cycles within the detector's sensitivity band.

%%%%%%%%%%%%%%%%%%%%%%%%%%%%%%%%%%%%%%%%%%%%%%%%%%%%%%%%%%%%%%%%%%%%%%%%%%%%
%%%%%%%%%%%%%%%%%%%%%%%%%%%%%%%%%%%%%%%%%%%%%%%%%%%%%%%%%%%%%%%%%%%%%%%%%%%%
\begingroup
\setlength{\tabcolsep}{0.75pt} 
\renewcommand{\arraystretch}{1.25} 
\begin{table*}[t]
\begin{center}
\begin{tabular}{|c|c|c|c|c|c|c|c|c|c|c|c|c|}
\hline
\multicolumn{1}{| c |}{ No.} & Event & Waveform & \multicolumn{5}{c|}{Properties of GR injections} & \multicolumn{2}{c|}{Properties of $\delta \hat{\phi}_{\rm PCA}^{(1)}$} & \multicolumn{2}{c|}{Properties of $\delta \hat{\phi}_{\rm PCA}^{(2)}$}\\[0.1cm]
\hline
 & & & $M$  &  $q$  & ($\chi_1,\chi_2$) & ($\theta_1$, $\theta_2$)  & SNR & Median and  & $\mathcal{Q}_{\rm GR}$ & Median and & $\mathcal{Q}_{\rm GR}$ \\[0.1cm]
 & & & ($M_{\odot}$) & & & & & 90\% errors & ($z_{\rm GR}$) & 90\% errors & ($z_{\rm GR}$)\\[0.2cm]
 \hline
 \hline
 1. & GW151226-like & \texttt{IMRPhenomXPHM} & 23.72 & 1.91 & (0.61, 0.455) & (1.02, 1.385) & 20 & ${-0.08}^{+0.18}_{-0.18}$ & 0.77 (0.7) & ${0.06}^{+0.83}_{-0.82}$ & 0.44 (0.13) \\[0.1cm]
    &     &    &    &     &       &    & 40 & $-0.02^{+0.08}_{-0.09}$ & 0.69 (0.46) & $-0.02^{+0.41}_{-0.4}$ & 0.53 (0.06) \\
\hline
 2. & GW190412-like & \texttt{IMRPhenomXHM} & 46.66 & 4.24 & (0.46,0.56) & (0,0) & 35 & $-0.03^{+0.14}_{-0.13}$ & 0.65 (0.32) & $0.11^{+0.75}_{-0.67}$ & 0.39 (0.26) \\[0.1cm]
    &      & \texttt{IMRPhenomXPHM} &    &     &     &  (0.68,1.66)  & 35 & $0.01^{+0.15}_{-0.12}$ & 0.43 (0.15) & $-0.03^{+0.86}_{-0.79}$ & 0.53 (0.06) \\

 3. & GW190814-like & \texttt{IMRPhenomXHM} & 27.2 & 9 & (0.03,0.47) & (0,0) & 25 & $-0.02^{+0.1}_{-0.09}$ & 0.64 (0.32) & $0.13^{+0.6}_{-0.6}$ & 0.35 (0.34) \\[0.1cm]
    &      & \texttt{IMRPhenomXPHM} &    &     &       &  (1.57,1.46)  & 25 & $0.0^{+0.08}_{-0.07}$ & 0.48 (0.04) & $0.08^{+0.56}_{-0.54}$ & 0.4 (0.23) \\
\hline
\hline
\end{tabular}
\caption{Summary of the {\tt PCA-TGR} results from injection studies using simulated GR-consistent signals from GW151226-like, GW190412-like, and GW190814-like BBH systems, conducted within the {\tt TIGER} framework, as discussed in Sec.~\ref{subsec:prec_HM_inj}.
Each row lists the waveform model used for injection (with recovery performed using its corresponding parametrized version, which is not separately listed), along with the binary parameters and the statistical properties of the leading two PCA parameters. Posterior distributions for the dominant two PCA components are shown in Fig.~\ref{fig:GR_inj_GW_events}.
}
\label{tab:table_gr_inj_gw_events}
\end{center}
\end{table*}
\endgroup
%%%%%%%%%%%%%%%%%%%%%%%%%%%%%%%%%%%%%%%%%%%%%%%%%%%%%%%%%%%%%%%%%%%%%%%%%%%%
%%%%%%%%%%%%%%%%%%%%%%%%%%%%%%%%%%%%%%%%%%%%%%%%%%%%%%%%%%%%%%%%%%%%%%%%%%%%

\subsubsection{Precessing systems} \label{subsec:prec_gr_injs}
We now present results for a set of GR-consistent injection runs involving precessing BBHs. 
The configurations span a range of effective precession parameters $\chi_p$~\cite{Hannam:2013oca,Schmidt:2014iyl} and network SNRs, allowing us to investigate the performance of the {\tt PCA-TGR} in recovering the injected GR signal under varying levels of precessional dynamics and signal strength.

All injections correspond to binary systems with total mass $30 M_{\odot}$ and mass ratio $q = 2$. 
To probe both moderately and strongly precessing regimes, we consider two configurations differing in spin magnitudes and orientations. 
For a high-precession configuration, we consider $(\chi_1, \chi_2) = (0.8, 0.7)$ with tilt angles $(\theta_1, \theta_2) = (\pi/3, \pi/4)$, corresponding to $\chi_p = 0.69$. 
For a low-precession configuration, we use $(\chi_1, \chi_2) = (0.3, 0.2)$ with smaller tilt angles $(\theta_1, \theta_2) = (\pi/9, \pi/4)$, yielding $\chi_p = 0.102$. In each case, we choose $\theta_{JN} = 2.7$ radians, $\phi_{JL} = 1.7$ radians, and $\phi_{12} = 3$ radians.
Each system is analyzed at two network SNRs: $\sim 20$ and $\sim 40$, corresponding to low and high signal strengths, respectively.

For GR-consistent injection, we employ the \texttt{IMRPhenomXP} waveform model \cite{Pratten:2020ceb}, a frequency-domain phenomenological waveform approximant that extends the \texttt{IMRPhenomXAS} \cite{Pratten:2020fqn} by incorporating spin-induced precession effects in the quadrupolar mode. 
For recovery in the parameter estimation process, we use the {\tt parametrized IMRPhenomXP}~\cite{Roy:2025gzv} waveform model approximant to produce discrete samples of the posterior probability distribution given in Eq.~(\ref{eq:Bayes_theorem}) within the {\tt TIGER} framework and subsequently perform the {\tt PCA-TGR} analysis.

The binary configurations used in this analysis are summarized in Table~\ref{tab:table_gr_inj_precess_spin}, which also reports the median and 90\% credible intervals of the posteriors for the leading two PCA parameters. Here, we find that the JS divergence between the posterior and prior distributions exceeds 0.1 bits only for the first two leading PCA parameters. The values of $\mathcal{Q}_{\rm GR}$ and $z_{\rm GR}$ for the PCA posteriors are also provided. Figure~\ref{fig:GR_inj_prec} displays the posterior distributions of $\delta\hat{\phi}^{(1)}_{\rm PCA}$ and $\delta\hat{\phi}^{(1)}_{\rm PCA}$ as violin plots for all injection configurations. 
We find that, in all cases, the GR value lies well within the 90\% credible regions of the PCA parameters. This is reflected in the corresponding $\mathcal{Q}_{\rm GR}$ and $z_{\rm GR}$ values reported in Table~\ref{tab:table_gr_inj_precess_spin}.
Moreover, as expected, the constraints on the PCA parameters tighten with increasing SNR, illustrating the dependence of the {\tt PCA-TGR} method’s constraining power on the signal strength.

%%%%%%%%%%%%%%%%%%%%%%%%%%%%%%%%%%%%%%%%%%%%%%%%%%%%%%%%%%%%%%%%%%%%%%%%%%%%
%%%%%%%%%%%%%%%%%%%%%%%%%%%%%%%%%%%%%%%%%%%%%%%%%%%%%%%%%%%%%%%%%%%%%%%%%%%%
\begin{figure*}[hbt!]
 \centering
    \includegraphics[width=0.98\linewidth]{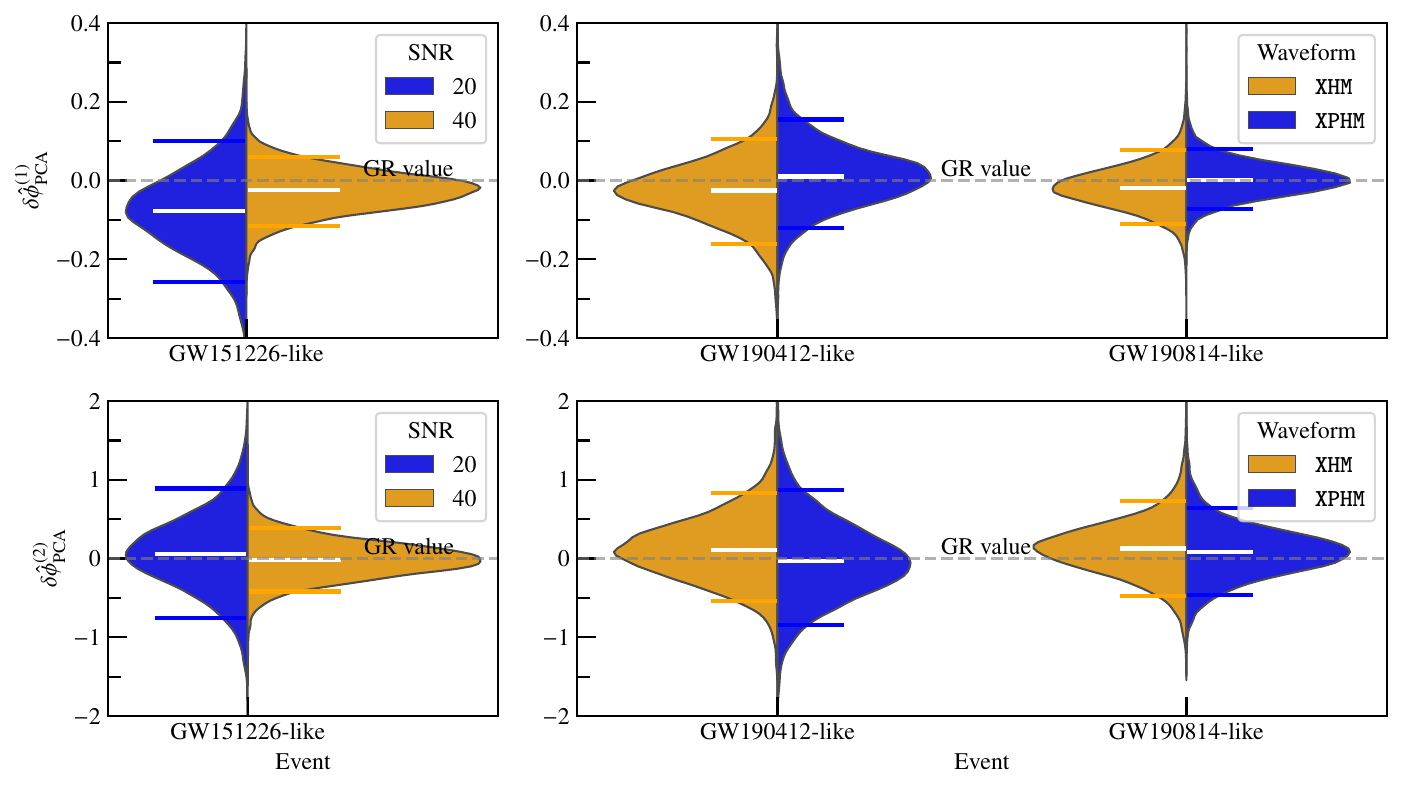}
    \caption{Posterior distributions of the dominant two PCA parameters ($\delta\hat{\phi}^{(1)}_{\rm PCA}$, top row; $\delta\hat{\phi}^{(2)}_{\rm PCA}$, bottom row) obtained from GR-consistent injections based on GW151226-, GW190412-, and GW190814-like BBH systems, as described in Sec.~\ref{subsec:prec_HM_inj}. The left panel compares PCA posteriors for GW151226-like injections at network SNRs of 20 and 40, using \texttt{IMRPhenomXPHM} for injection and its parametrized version for recovery.
    The right panel compares aligned-spin (\texttt{IMRPhenomXHM} for injection and its parametrized version for recovery) and precessing (\texttt{IMRPhenomXPHM} for injection and its parametrized version for recovery) injections for GW190412-like and GW190814-like systems. The colored and white horizontal bars have the same meaning as in Fig.~\ref{fig:gr_inj_aligned_spin}. In all cases, the PCA constraints are consistent with the GR value of 0, marked by the dashed gray line. 
    }
    \label{fig:GR_inj_GW_events}
\end{figure*}
%%%%%%%%%%%%%%%%%%%%%%%%%%%%%%%%%%%%%%%%%%%%%%%%%%%%%%%%%%%%%%%%%%%%%%%%%%%%
%%%%%%%%%%%%%%%%%%%%%%%%%%%%%%%%%%%%%%%%%%%%%%%%%%%%%%%%%%%%%%%%%%%%%%%%%%%%

\subsubsection{Systems with precession and higher-order harmonics}
\label{subsec:prec_HM_inj}
We now extend our analysis to investigate the impact of higher-order harmonics and spin-induced precession by performing injection studies using binary parameters of known detected BBH systems. Specifically, we consider configurations inspired by three well-studied BBH merger events: GW151226 \cite{LIGOScientific:2016sjg}, GW190412~\cite{LIGOScientific:2020stg}, and GW190814~\cite{LIGOScientific:2020zkf}.
GW190412 and GW190814 exhibit pronounced mass asymmetry, which enhances the amplitude of higher-order harmonics in the GW signal. In contrast, GW151226 is more symmetric in masses and features moderate spin-induced precession. The injection parameters—component masses, spin magnitudes, and tilt angles—are selected to approximate the inferred properties of these observed systems and are summarized in Table~\ref{tab:table_gr_inj_gw_events}.
 
To systematically disentangle the effects of precession and higher modes, we perform injections and recoveries using two types of waveform models. The first, \texttt{IMRPhenomXHM}~\cite{Garcia-Quiros:2020qpx} (for injection) and its parametrized version (for recovery), includes higher-order modes but assumes spins aligned with the orbital angular momentum. The second, \texttt{IMRPhenomXPHM} \cite{Pratten:2020ceb} (for injection) and its parametrized version (for recovery), extends the former by incorporating spin-precession effects.

For GW151226-like injections (mass ratio $q \sim 2$, $\chi_p \sim 0.52$), we use \texttt{IMRPhenomXPHM} for injection and \texttt{parametrized IMRPhenomXPHM} for recovery, with network SNRs of 20 and 40.
For the mass-asymmetric GW190412-like ($q \sim 4$, $\chi_p \sim 0.29$) and GW190814-like ($q \sim 9$, $\chi_p \sim 0.07$) systems, we conduct separate injection studies: one set using \texttt{IMRPhenomXHM} with zero spin tilts (aligned-spin case) and its parametrized version for recovery, and another set using \texttt{IMRPhenomXPHM} along with its parametrized version to include spin precession effects.
This strategy allows for a direct comparison of aligned and precessing versions of the same system, thereby isolating the contributions from spin precession and higher-order modes. Such comparisons are essential, as spin-induced precession can alter the mixing of angular modes, thereby modulating the detectability and influence of higher-order harmonics on the waveform. 
Verifying that PCA-based tests remain robust in these scenarios is crucial for assessing {\tt PCA-TGR}'s applicability to future complex signals.

For GW190814-like injections, we use the noise PSDs estimated with {\tt BAYESWAVE}~\cite{Pankow:2018qpo,Cornish:2020dwh} from the actual event data for each detector~\cite{LIGOScientific:2020zkf}. For the other two events, we use the same detector-specific noise PSDs as in the previous sections. The results of this injection analysis are summarized in Table~\ref{tab:table_gr_inj_gw_events} and Fig.~\ref{fig:GR_inj_GW_events}. Here, we also find that the JS divergence between the posterior and prior distributions exceeds 0.1 bits for the first two leading PCA parameters in all cases while it is marginally greater than 0.1 bits for the third leading PCA parameter for some cases. The statistics of the two leading PCA posteriors are listed in Table~\ref{tab:table_gr_inj_gw_events}. Figure~\ref{fig:GR_inj_GW_events} shows their posterior distributions. The top row corresponds to $\delta\hat{\phi}^{(1)}_{\rm PCA}$ and the bottom row to $\delta\hat{\phi}^{(2)}_{\rm PCA}$. In each row, the left panel shows the posteriors of GW151226-like injections at low and high SNR. The right panel shows the posteriors for GW190412-like and GW190814-like injections, considering both aligned-spin and precessing cases. In summary, the PCA posteriors are consistent with GR in all cases, with the injected value of zero lying within the 90\% credible intervals. The bounds on the PCA parameters for the GW190814-like injection are relatively tighter compared to the others, due to the greater strength of higher-order modes in the injected signal, which helps break degeneracies between different model parameters.

\subsection{FTI framework}
\label{sec:FTI_inj}

\begingroup
\setlength{\tabcolsep}{0.75pt} % Default value: 6pt
\renewcommand{\arraystretch}{1.25} % Default value: 1
\begin{table*}
\begin{center}
\begin{tabular}{|c|c|c|c|c|c|c|c|}
\hline
\multicolumn{1}{| c |}{ No.} & \multicolumn{3}{c|}{} & \multicolumn{2}{c|}{Properties of $\delta \hat{\phi}_{\rm PCA}^{(1)}$} & \multicolumn{2}{c|}{Properties of $\delta \hat{\phi}_{\rm PCA}^{(2)}$}\\[0.2cm]
\hline
 & Event & Waveform injection &  SNR & Median and  & $\mathcal{Q}_{\rm GR}$ & Median and & $\mathcal{Q}_{\rm GR}$ \\[0.2cm]
 &  & Waveform recovery &  & 90\% errors & ($z_{\rm GR}$) & 90\% errors & ($z_{\rm GR}$)\\[0.2cm]
 \hline
 \hline
 1. & GW150914-like & {\tt SEOBNRv4$\textunderscore$ROM} & $20$ & $0.03_{-0.13}^{+0.22}$ & 0.39 (0.23) & $-0.04_{-0.65}^{+0.55}$ & 0.55 (0.12) \\[0.2cm]
    &    &   {\tt parametrized SEOBNRv4$\textunderscore$ROM}  &           $ 40$ & $0.02_{-0.08}^{+0.15}$ & 0.35 (0.31) & $0.01_{-0.32}^{+0.35}$ & 0.49 (0.04) \\
    \hline
 2. & GW150914-like  & {\tt SEOBNRv4HM$\textunderscore$ROM} & $20$ & $0.02_{-0.08}^{+0.15}$ & 0.38 (0.26) & $-0.03_{-0.61}^{0.53}$& 0.53 (0.08) \\ [0.2cm]
  & & {\tt parametrized SEOBNRv4HM$\textunderscore$ROM} & $ 40$ & $0.01_{-0.06}^{+0.1}$ & 0.38 (0.26) & $0.02_{-0.28}^{+0.32}$ &0.46 (0.09) \\
  \hline
\multicolumn{8}{| c |}{Effect of precession on PCA}\\
\hline
3. & $M =30  M_{\odot}, q=2,\, \chi_p = 0.69$  & {\tt IMRPhenomXPHM} & $20$ & $-0.07_{-0.15}^{+0.02}$ & 0.91 (1.31) & $-0.62_{-1.04}^{-0.26}$& 1.0 (2.59) \\ [0.2cm]
& & {\tt parametrized SEOBNRv4HM$\textunderscore$ROM} & & & & &\\
\hline
4. & $M =30  M_{\odot}, q=2,\, \chi_p = 0.102$  & {\tt IMRPhenomXPHM} & $20$ & $-0.06_{-0.15}^{+ 0.04}$ & 0.86 (0.93) & $-0.23_{-0.68}^{+0.15}$& 0.83 (0.9) \\ [0.2cm]
& & {\tt parametrized SEOBNRv4HM$\textunderscore$ROM } & & & & &\\
\hline
\end{tabular}
\caption{Summary of the {\tt PCA-TGR} results from the simulated BBH GR injection study conducted within the {\tt FTI} framework, as described in Sec.~\ref{sec:FTI_inj}. Each row lists the binary parameters, the waveform models used for injection and recovery, and the statistical properties of the leading two PCA parameters. Posterior distributions for the leading two PCA components are shown in Figs.~\ref{fig:FTI_GR_injection} and~\ref{fig:FTI_precession} corresponding to the first two and the last two entries in the table, respectively.
}
\label{tab:table_gr_inj_fti}
\end{center}
\end{table*}
\endgroup

%%%%%%%%%%%%%%%%%%%%%%%%%%%%%%%%%%%%%%%%%%%%%%%%%%%%%%%%%%%%%%%%%%%%
%%%%%%%%%%%%%%%%%%%%%%%%%%%%%%%%%%%%%%%%%%%%%%%%%%%%%%%%%%%%%%%%%%%%
\begin{figure*}[hbt!]
    \centering
\includegraphics[width=0.4955\textwidth]{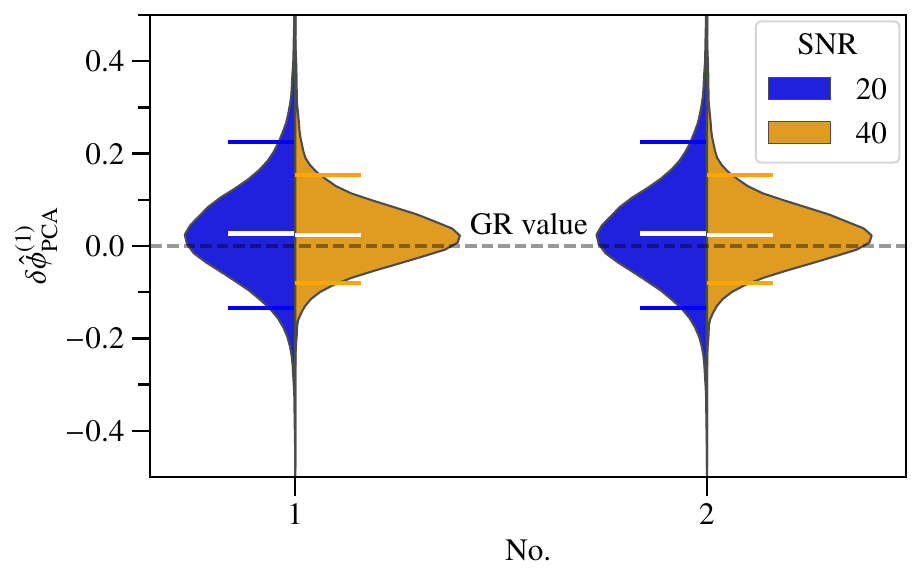}
\includegraphics[width=0.4955\textwidth]{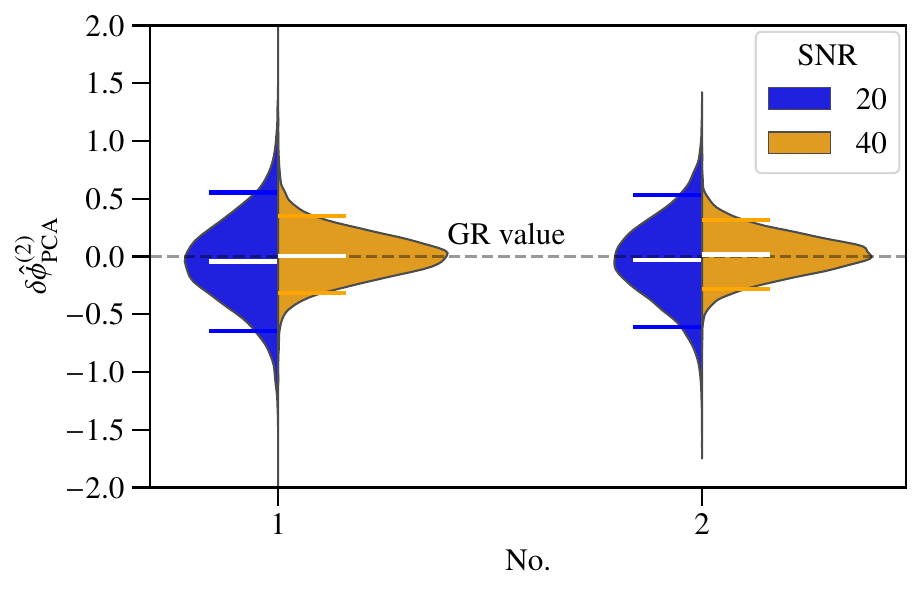}
    \caption{Violin plots showing the posteriors of $\delta \hat{\phi}^{(1)}_{\rm PCA}$ and $\delta \hat{\phi}^{(2)}_{\rm PCA}$ from the GR-consistent aligned-spin injection analysis within the {\tt FTI} framework, using the {\tt SEOBNRv4$\textunderscore$ROM} (x-label:1) and {\tt SEOBNRv4HM$\textunderscore$ROM} (x-label:2) waveform models for GW150914-like BBH systems, as listed in entries 1 and 2 of Table~\ref{tab:table_gr_inj_fti} and discussed in Sec.~\ref{sec:FTI_inj}. The horizontal bars have the same meaning as in Fig.~\ref{fig:gr_inj_aligned_spin}. In every case, the PCA posteriors remain consistent with the GR prediction of zero, indicated by the dashed gray line.
    }
    \label{fig:FTI_GR_injection}
\end{figure*}
%%%%%%%%%%%%%%%%%%%%%%%%%%%%%%%%%%%%%%%%%%%%%%%%%%%%%%%%%%%%%%%%%%%%
%%%%%%%%%%%%%%%%%%%%%%%%%%%%%%%%%%%%%%%%%%%%%%%%%%%%%%%%%%%%%%%%%%%%

%%%%%%%%%%%%%%%%%%%%%%%%%%%%%%%%%%%%%%%%%%%%%%%%%%%%%%%%%%%%%%%%%%%%
%%%%%%%%%%%%%%%%%%%%%%%%%%%%%%%%%%%%%%%%%%%%%%%%%%%%%%%%%%%%%%%%%%%%
\begin{figure*}[hbt!]
    \centering
\includegraphics[width=0.4955\textwidth]{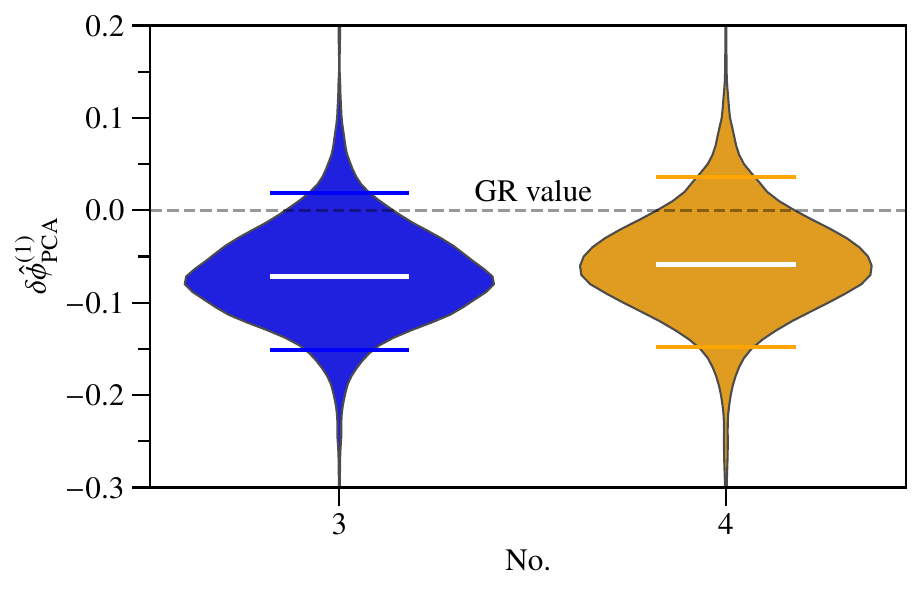}
\includegraphics[width=0.4955\textwidth]{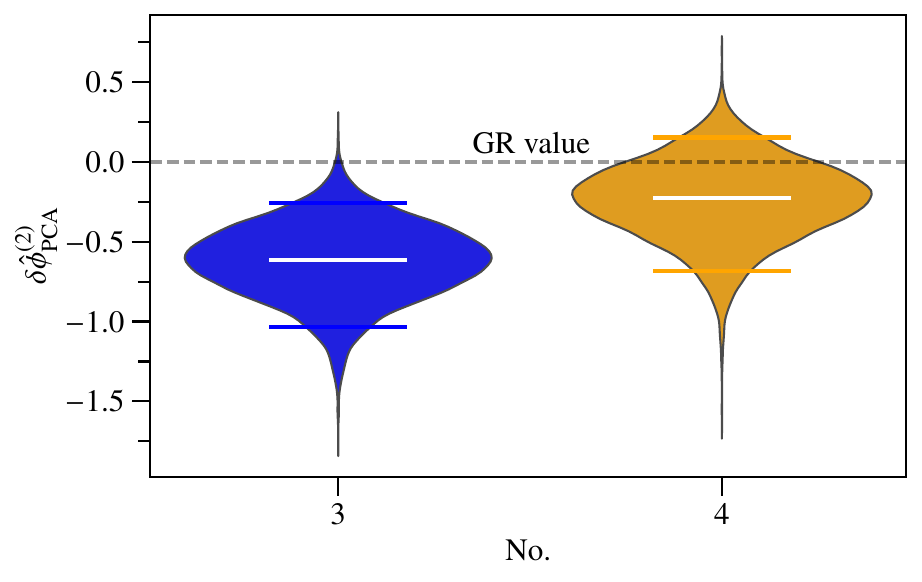}
    \caption{Violin plots showing the posteriors of $\delta \hat{\phi}^{(1)}_{\rm PCA}$ and $\delta \hat{\phi}^{(2)}_{\rm PCA}$ from the GR-consistent precessing-spin injection analysis within the {\tt FTI} framework, as presented in entries 3 and 4 of Table~\ref{tab:table_gr_inj_fti} and discussed in Sec.~\ref{sec:FTI_inj}. GR-consistent GW signals are simulated using the {\tt IMRPhenomXPHM} waveform model, and parameter estimation is performed using the {\tt parametrized SEOBNRv4HM$\textunderscore$ROM} model. The horizontal bars have the same meaning as in Fig.~\ref{fig:gr_inj_aligned_spin}. The median values of the PCA parameter posteriors show notable deviations from zero in all cases. Specifically, for the strongly precessing binary configuration, the GR value is excluded from the 90\% credible interval of the $\delta \hat{\phi}^{(2)}_{\rm PCA}$ posterior. 
    }
    \label{fig:FTI_precession}
\end{figure*}
%%%%%%%%%%%%%%%%%%%%%%%%%%%%%%%%%%%%%%%%%%%%%%%%%%%%%%%%%%%%%%%%
%%%%%%%%%%%%%%%%%%%%%%%%%%%%%%%%%%%%%%%%%%%%%%%%%%%%%%%%%%%%%%%%%%%%

In the previous subsection, we studied the efficacy of {\tt PCA-TGR} within the {\tt TIGER} framework using a variety of simulated GR-consistent injected signals. We now shift out focus to investigate the efficacy of {\tt PCA-TGR} within the {\tt FTI} framework. One important point to note is that, in its current implementation, the {\tt FTI} framework uses aligned-spin waveform models.

We consider GW150914-like BBH systems, with injected binary parameters set to the median values obtained from the parameter estimation of the GW150914 event~\cite{GW150914PEup}. The luminosity distance is varied to achieve the desired network SNR values of 20 (low) and 40 (high). To simulate GR-consistent GW signals, we first use the {\tt SEOBNRv4$\textunderscore$ROM} waveform approximant~\cite{Bohe:2016gbl}. For parameter estimation, we employ the {\tt parametrized SEOBNRv4$\textunderscore$ROM}~\cite{FTI,Sanger:2024axs} model for signal recovery, producing discrete samples from the posterior probability distribution given in Eq.~(\ref{eq:Bayes_theorem}), within the {\tt FTI} framework. We repeat the same exercise but now using the {\tt SEOBNRv4HM$\textunderscore$ROM} waveform model~\cite{Cotesta:2018fcv,Cotesta:2020qhw}, which incorporates higher-order modes beyond the quadrupolar ones. 

The results from this analysis are summarized in Table~\ref{tab:table_gr_inj_fti} (first two entries) and Fig.~\ref{fig:FTI_GR_injection}. The JS divergence between the posterior and prior distributions exceeds 0.1 bits for the first three leading PCA parameters in all cases; however, we present results only for the first two.
Similarly to the previous subsection, the statistical properties of the two leading PCA parameters are listed in Table~\ref{tab:table_gr_inj_fti}, while their posterior distributions are shown as violin plots in Fig.~\ref{fig:FTI_GR_injection}. In every case, the PCA posteriors are found to be consistent with GR, with the injected value of zero lying within the 90\% credible intervals. The {\tt PCA-TGR} analysis with similar GR injections has also been repeated using the {\tt SEOBNRv5\_ROM} waveform family~\cite{Pompili:2023tna}, yielding results similar to those from {\tt SEOBNRv4\_ROM}, although these results are not included here for brevity.

Within the {\tt FTI} framework, we also explore the effect of spin precession on the {\tt PCA-TGR} analysis since the waveform models used in the {\tt FTI} analysis do not account for it. 
We consider the BBH system with a total mass of $30 M_{\odot}$, a mass ratio of $q=2$, and a network SNR of approximately 20. 
We study this system using two different values of the effective spin-precession parameter, $\chi_p = 0.102$ (weak precession) and $\chi_p = 0.69$ (strong precession), obtained by varying the spin magnitudes and tilt angles. For each case, GR-consistent injected signals are simulated using the {\tt IMRPhenomXPHM} waveform model, which incorporates both higher-order modes and precession effects. Parameter estimation is performed using the {\tt parametrized SEOBNRv4HM\_ROM} model, which accounts for higher-order modes but not for precession.

Similar to the previous injection analysis, we again find that the JS divergence between the posterior and prior distributions exceeds 0.1 bits for the three leading PCA parameters in both injections. The statistical properties of the first two PCA parameters for both cases are shown in entries 3 and 4 of Table~\ref{tab:table_gr_inj_fti}, respectively.  Since this is a GR injection, any apparent GR deviation observed in the posteriors of the PCA parameters can be attributed to the effects of unaccounted spin precession in the waveform model used for parameter estimation during recovery. As shown in Fig.~\ref{fig:FTI_precession}, the posterior medians of the leading two PCA parameters exhibit significant offsets from zero in both cases. In particular, the GR value is excluded at 90\% confidence in the posterior of $\delta \hat{\phi}^{(2)}_{\rm PCA}$ for the strongly precessing binary configuration. 
The posteriors of $\delta \hat{\phi}^{(3)}_{\rm PCA}$ are almost an order of magnitude less constrained than those of $\delta \hat{\phi}^{(2)}_{\rm PCA}$ and are not shown here. Although their posteriors are consistent with GR, the posterior medians show a slight offset from zero. The conclusion from this injection study is that using an aligned-spin waveform model in the {\tt FTI} framework could lead to a false violation of GR in the {\tt PCA-TGR} analysis if significant spin precession is present in the signal.

\section{ Response of PCA-TGR to Non-GR Injections}\label{sec:non-GRinj} 
After demonstrating the ability of {\tt PCA-TGR} to recover GR in an unbiased manner when the signal in the data is that of a BBH merger in GR---with the exception of its response in the {\tt FTI} framework when an aligned-spin waveform is used to analyze strongly precessing signals---we now turn our attention to non-GR injections and {\tt PCA-TGR}'s ability to detect GR deviations.

\subsection{TIGER framework}

%%%%%%%%%%%%%%%%%%%%%%%%%%%%%%%%%%%%%%%%%%%%%%
%%%%%%%%%%%%%%%%%%%%%%%%%%%%%%%%%%%%%%%%%%%
%\pagebreak
\begingroup
\setlength{\tabcolsep}{0.5pt} 
\renewcommand{\arraystretch}{1.25}
\begin{table*}[hbt!]
\begin{center}
\begin{tabular}{|c|c|c|c|c|c|c|c|c|c|c|}
\hline
\multicolumn{1}{| c |}{ No.} & \multicolumn{2}{c|}{Properties of non-GR injections} & \multicolumn{4}{c|}{Properties of $\delta \hat{\phi}_{\rm PCA}^{(1)}$} & \multicolumn{4}{c|}{Properties of $\delta \hat{\phi}_{\rm PCA}^{(2)}$}\\
\cline{2-11}
\multirow{2}{*}{} & \multirow{2}{*}{$\{\delta\hat{\phi}_{k}\}$} & \multicolumn{1}{c|}{} & \multicolumn{2}{c|}{Median and} & \multicolumn{2}{c|}{$\mathcal{Q}_{\rm GR}$} & \multicolumn{2}{c|}{Median and} & \multicolumn{2}{c|}{$\mathcal{Q}_{\rm GR}$}\\
& & \multicolumn{1}{c|}{SNR} & \multicolumn{2}{c|}{90\% errors} & \multicolumn{2}{c|}{($z_{\rm GR}$)} & \multicolumn{2}{c|}{90\% errors} & \multicolumn{2}{c|}{($z_{\rm GR}$)}\\
\cline{4-11}
& & & GW1509 & GW1512 & GW1509 & GW1512 & GW1509 & GW1512 & GW1509 & GW1512\\
& & & 14-like & 26-like & 14-like & 26-like & 14-like & 26-like & 14-like & 26-like\\
\hline
1. & $\delta\hat{\phi}_{a}=0\, (a=-2,0,1,2)$ & 20 & $0.10^{+0.25}_{-0.26}$ & $0.00^{+0.20}_{-0.23}$ & 0.26 (0.62) & 0.50 (0.01) & $-0.07^{+1.71}_{-2.04}$ & $0.07^{+0.84}_{-1.00}$ & 0.53 (0.06) & 0.45 (0.11)\\
& $\delta\hat{\phi}_{b}=0.1\, (b=3,4,$ & & & & & & & & &\\
& $5l,6,6l,7)$  & 40 & $0.11^{+0.17}_{-0.14}$ & $0.06^{+0.11}_{-0.11}$ & 0.09 (1.17) & 0.16 (0.94) & $-0.07^{+0.98}_{-1.15}$ & $0.03^{+0.41}_{-0.44}$ & 0.55 (0.11) & 0.45 (0.10)\\
\hline
2. & $\delta\hat{\phi}_{a}=0\, (a=-2,0,1,2)$ & 20 & $0.51^{+0.19}_{-0.19}$ & $0.41^{+0.24}_{-0.24}$ & 0.00 (4.09) & 0.01 (2.74) &$0.03^{+2.16}_{-2.05}$ & $0.54^{+0.96}_{-0.84}$ & 0.49 (0.02) & 0.15 (0.98)\\
& $\delta\hat{\phi}_{b}=0.5\, (b=3,4,$ & & & & & & & & &\\
& $5l,6,6l,7)$  & 40 & $0.52^{+0.13}_{-0.12}$ & $0.48^{+0.14}_{-0.14}$ & 0.00 (6.50) & 0.00 (5.66) & $-0.17^{+1.16}_{-1.23}$ & $0.15^{+0.53}_{-0.46}$ & 0.59 (0.23) & 0.30 (0.49)\\
\hline
3. & $\delta\hat{\phi}_{a}=0\, (a=-2,0,1,2)$ & 20 & $0.92^{+0.15}_{-0.16}$  & $0.90^{+0.20}_{-0.16}$ & 0.00 (9.16) & 0.00 (7.90) & $-0.03^{+2.89}_{-2.67}$ & $0.48^{+1.11}_{-0.95}$ & 0.51 (0.02) & 0.21 (0.75)\\
& $\delta\hat{\phi}_{b}=0.9\, (b=3,4,$ & & & & & & & & &\\
& $5l,6,6l,7)$  & 40 & $0.93^{+0.12}_{-0.10}$ & $0.93^{+0.12}_{-0.10}$ & 0.00 (12.83) & 0.00 (13.17) & $-0.31^{+1.49}_{-1.64}$ & $0.12^{+0.57}_{-0.53}$ & 0.63 (0.33) & 0.35 (0.37)\\
\hline
4. & $\delta\hat{\phi}_{a}=0\, (a=-2,0,1,2)$ & 20 & $0.35^{+0.21}_{-0.23}$  & $0.18^{+0.25}_{-0.26}$ & 0.01 (2.48) & 0.12 (1.16) & $0.09^{+1.89}_{-1.97}$ & $0.48^{+0.93}_{-0.77}$ & 0.47 (0.08) & 0.15 (0.89)\\
  & $\delta\hat{\phi}_{3}=0.3,\, \delta\hat{\phi}_{4}=0.4$ & & & & & & & & &\\
  & $\delta\hat{\phi}_{5l}=0.5,\,\delta\hat{\phi}_{6}=0.6$ & & & & & & & & &\\
& $\delta\hat{\phi}_{6l}=0.7,\,\delta\hat{\phi}_{7}=0.8$ & 40 & $0.37^{+0.14}_{-0.14}$ & $0.30^{+0.13}_{-0.14}$ & 0.00 (4.13) & 0.00 (3.58) & $-0.04^{+1.07}_{-1.15}$ & $0.28^{+0.45}_{-0.39}$ & 0.52 (0.05) & 0.12 (1.06)\\
\hline
5. & $\delta\hat{\phi}_{a}=0\, (a=-2,0,1,2)$ & 20 & $0.77^{+0.17}_{-0.16}$  & $0.73^{+0.21}_{-0.17}$ & 0.00 (7.17) & 0.00 (6.17) & $-0.05^{+2.64}_{-2.33}$ & $0.43^{+1.17}_{-0.95}$ & 0.51 (0.03) & 0.23 (0.67)\\
  & $\delta\hat{\phi}_{3}=0.8,\, \delta\hat{\phi}_{4}=0.7$ & & & & & & & & &\\
  & $\delta\hat{\phi}_{5l}=0.6,\,\delta\hat{\phi}_{6}=0.5$ & & & & & & & & &\\
& $\delta\hat{\phi}_{6l}=0.4,\, \delta\hat{\phi}_{7}=0.3$ & 40 & $0.78^{+0.11}_{-0.10}$ & $0.78^{+0.14}_{-0.11}$ & 0.00 (11.03) & 0.00 (10.43) & $-0.37^{+1.35}_{-1.41}$ & $0.07^{+0.60}_{-0.52}$ & 0.67 (0.44) & 0.42 (0.21)\\
\hline
6. & $\delta\hat{\phi}_{a}=0\, (a=-2,0,1,2)$ & 20 & $-0.50^{+0.30}_{-0.27}$ & $-0.56^{+0.21}_{-0.20}$ & 0.99 (2.83) & 1.00 (4.33) & $0.00^{+1.86}_{-2.42}$ & $0.00^{+0.90}_{-1.03}$ & 0.50 (0.00) & 0.50 (0.00)\\
& $\delta\hat{\phi}_{b}=-0.5\, (b=3,4,$ & & & & & & & & &\\
& $5l,6,6l,7)$  & 40 & $-0.49^{+0.21}_{-0.17}$ & $-0.53_{-0.11}^{+0.13}$ & 1.00 (4.11) & 1.00 (7.59) & $-0.03^{+1.15}_{-1.56}$ & $-0.11^{+0.51}_{-0.52}$ & 0.52 (0.04) & 0.65 (0.34)\\
\hline
\hline
\end{tabular}
\caption{Summary of results from the simulated non-GR injection study within the {\tt TIGER} framework that demonstrates the capability of the {\tt PCA-TGR} method to detect GR violations in the PN phasing coefficients. The values of the fractional PN deviation parameter, binary systems, and the network SNRs (rounded to the nearest integer) of the injected signals are listed. For each non-GR injection, the table shows the median value, as well as the upper and lower limits of the 90\% credibility interval, $\mathcal{Q}_{\rm GR}$, and $z_{\rm GR}$ for the posteriors of $\delta\hat{\phi}^{(1)}_{\rm PCA}$ and $\delta \hat{\phi}^{(2)}_{\rm PCA}$. The posteriors on $\delta\hat{\phi}^{(1)}_{\rm PCA}$ exclude GR at $>3 \, \sigma$ in most of the cases, while the posteriors on $\delta\hat{\phi}^{(2)}_{\rm PCA}$ encompass the GR value within 90\% credibility in all cases. The {\tt PCA-TGR} method can rule out GR at $>3 \, \sigma$ for a fractional deviation of magnitude $>0.5$ in the PN phasing coefficients. 
}
\label{tab:table_non-GR_inj}
\end{center}
\end{table*}
\endgroup
%%%%%%%%%%%%%%%%%%%%%%%%%%%%%%%%%%%%%%%%%%%
%%%%%%%%%%%%%%%%%%%%%%%%%%%%%%%%%%%%%%%%%%%%%%

%%%%%%%%%%%%%%%%%%%%%%%%%%%%%%%%%%%%%%%%%%%%%%%%%
%%%%%%%%%%%%%%%%%%%%%%%%%%%%%%%%%%%%%%%%%%%%%%%%%%%%%%%%%%%%%%%%%%%%
\begin{figure*}[hbt!]
    \centering
\includegraphics[width=0.4955\textwidth]{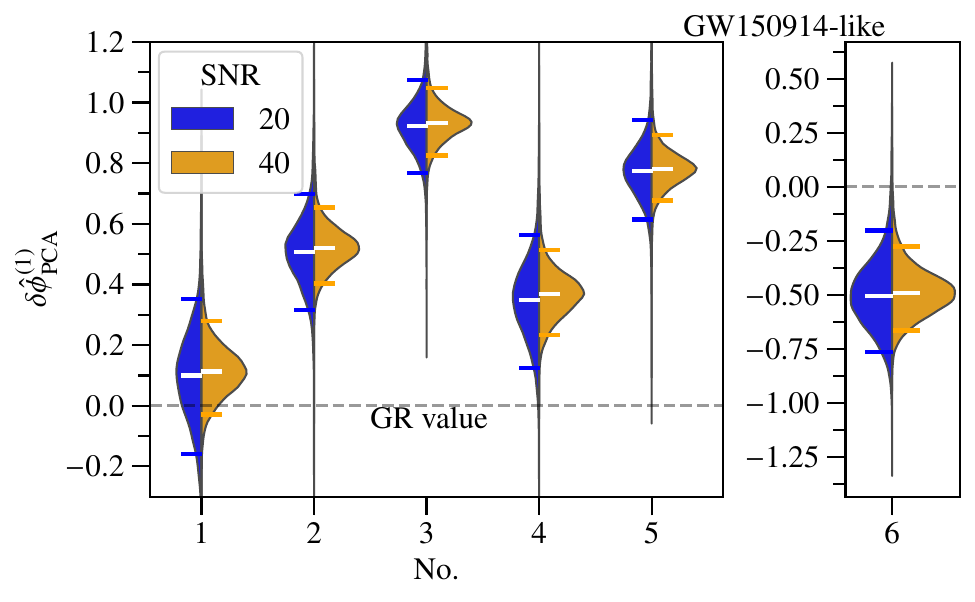}
\includegraphics[width=0.4955\textwidth]{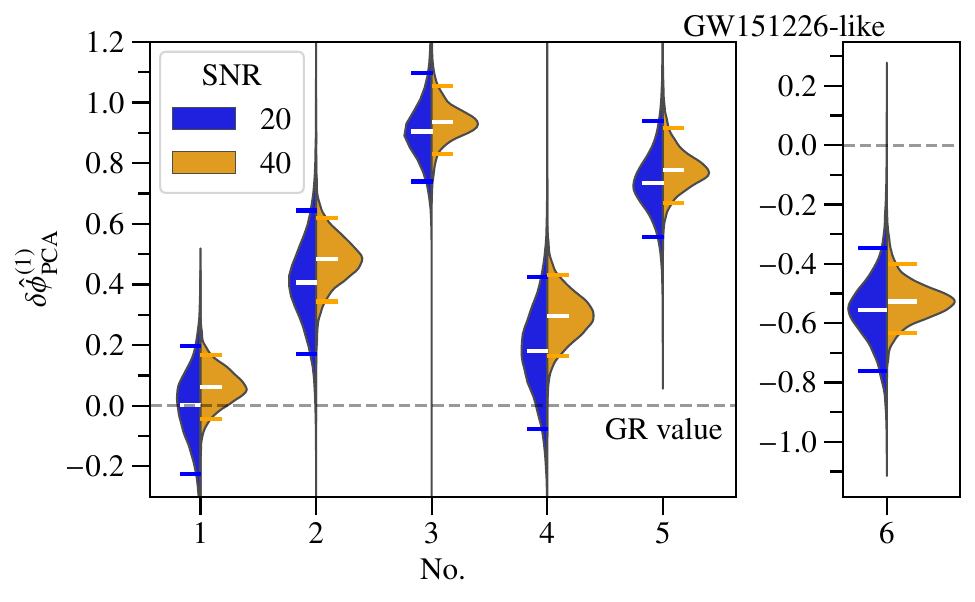}\\
\includegraphics[width=0.4955\textwidth]{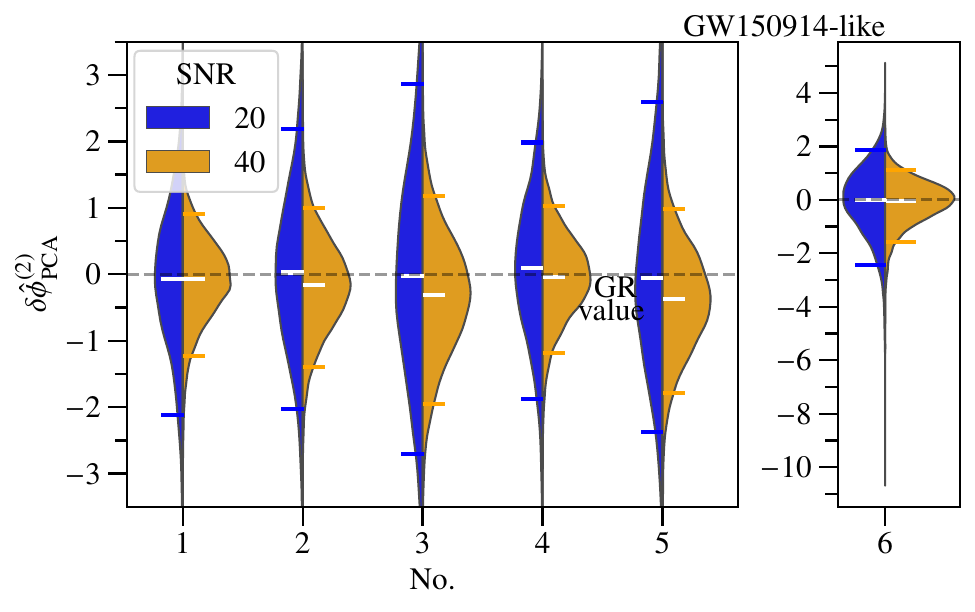}
\includegraphics[width=0.4955\textwidth]{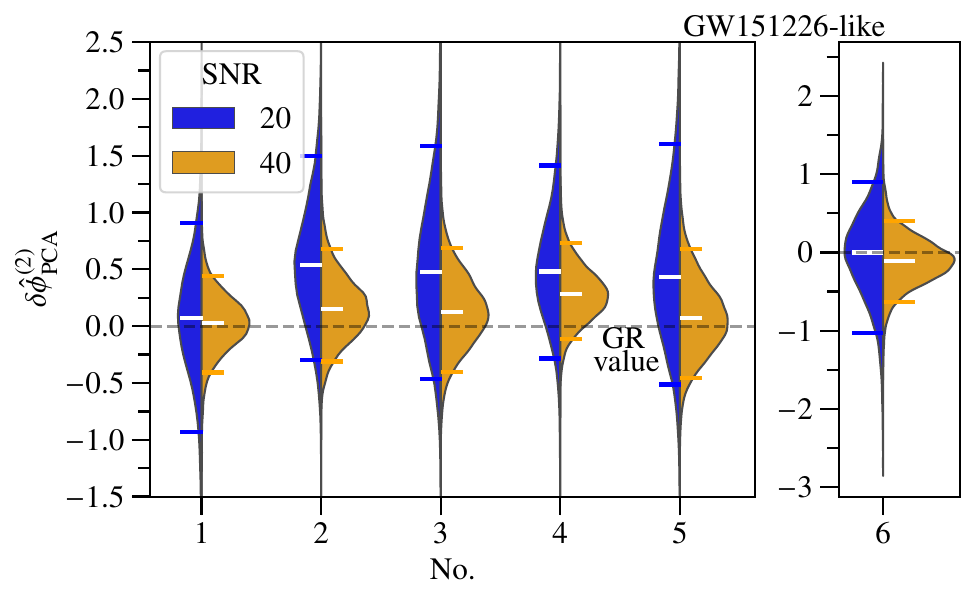}
    \caption{Posteriors of $\delta\hat{\phi}^{(1)}_{\rm PCA}$ (upper panel) and $\delta \hat{\phi}^{(2)}_{\rm PCA}$ (lower panel) for each simulated non-GR injection (the left and right panels show the results obtained from GW150914- and GW151226-like BBH injections, respectively) listed in Table~\ref{tab:table_non-GR_inj} presented as violin plots. The results from the injections with network SNR $\sim 20$ ($\sim 40$) are shown in blue (orange). The horizontal bars have the same meaning as in Fig.~\ref{fig:gr_inj_aligned_spin}. We mark the GR value of zero with dashed gray lines. 
    }
    \label{fig:non-GR}
\end{figure*}
%%%%%%%%%%%%%%%%%%%%%%%%%%%%%%%%%%%%%%%%%%%%%%%%%%%%%%%%%%%%%%%%

In order to simulate GR violating signals, we consider BBH waveforms whose inspiral phasing coefficients are deformed at different PN order using the {\tt TIGER} framework. We consider different types of non-GR injections which are summarized in Table \ref{tab:table_non-GR_inj}. Broadly speaking, the injections cover the following three scenarios:
\begin{enumerate}
    \item All the PN coefficients above 1PN are deformed by the same magnitude. We consider small, moderate, and large deviations which correspond to fractional deviations of 0.1, 0.5, and 0.9, respectively.
    \item The fractional deviations in the PN coefficients above 1PN increase (decrease) with PN order, ranging from 0.3 to 0.8 in steps of 0.1 (or vice versa).
    \item All the PN coefficients above 1PN are deformed by the same magnitude but by $-0.5$ to study the difference between a positive fractional deviation and a negative one.
\end{enumerate}
We consider GW150914- and GW151226-like BBH systems, using the same binary parameters as in the previous section. We take two different values of the luminosity distance for each of the chosen BBH systems to produce two different network SNRs of $\sim$20 and $\sim$40. 
We use the {\tt parametrized IMRPhenomXP} waveform model to generate the injected GW signal with GR violations and to perform parameter estimation while analyzing the injected signal in the {\tt TIGER} framework.
Indeed, the injected GR violations are \emph{ad hoc} and do not follow the predictions of any known modified theories of gravity. Our aim here is to assess the capability of the {\tt PCA-TGR} to detect a violation of GR when the GW signal does not follow the GR prediction.

Our results are summarized in Table \ref{tab:table_non-GR_inj} and Fig.~\ref{fig:non-GR}. We find that, in all injections, the JS divergence between the posterior and prior distributions exceeds 0.1 bits only for the first two leading PCA parameters.
Figure~\ref{fig:non-GR} shows the violin plots for the posterior probability distributions of the leading two PCA parameters from our injection analysis  using the {\tt TIGER} framework.
From Table~\ref{tab:table_non-GR_inj} and Fig.~\ref{fig:non-GR}, we find that the posteriors of $\delta\hat{\phi}^{(1)}_{\rm PCA}$ exclude GR at $>3\,\sigma$ level in most of the cases when we consider a fractional deviation of magnitude $>0.1$ in the PN phasing coefficients and for all cases with ${\rm SNR}$ of $40$.
However, for a fractional deviation of magnitude 0.1, the {\tt PCA-TGR} is unable to detect GR violation with a high confidence level even for signals with a network SNR of $40$. Our simulations currently are not exhaustive enough to find the SNR at which a fractional deviation of 0.1 at every PN order will show up as a GR violation in the PCA parameters. However, a simple scaling suggests that a minimum SNR of $\sim$103 (126) is required to detect this level of GR violation at more than the $3\,\sigma$ level for a GW150914-like (GW151226-like) system. 
In all the simulations considered here, the subleading PCA parameters show consistency with GR within 90\% credibility. 
For deviations that are fractional in nature, such as those considered here, $\delta\hat{\phi}^{(1)}_{\rm PCA}$ is most efficient in detecting a GR violation. However, we cannot comment whether this is a generic feature for all possible types of GR violations with our limited set of injections. As these deviations are not motivated by any specific theory of gravity but are representative examples for demonstration of the method, we do not consider more combinations of the fractional deformation values. We will discuss other forms of GR deviations and the response of the {\tt PCA-TGR} in the next section.

In most of the non-GR injections, the error bars on the first PCA parameter $\delta\hat{\phi}^{(1)}_{\rm PCA}$ are comparable for GW150914-like and GW151226-like systems, with GW150914-like injections providing slightly better constraints in most cases. In contrast, GW151226-like systems consistently provide better constraints on the second PCA parameter $\delta\hat{\phi}^{(2)}_{\rm PCA}$.
For $\delta\hat{\phi}^{(1)}_{\rm PCA}$, GW150914-like injections generally yield higher $z_{\rm GR}$ values compared to GW151226-like injections, except for injection No. 6 and the SNR 40 case in injection No. 3.
This is likely due to the trade-off between the relative weights of the different $\delta\hat{\phi}_{b}$ in the PCA parameters, the magnitudes of these deformations, and the measurement uncertainties associated with these deviations. For detector configurations with better low-frequency sensitivity, GW151226-like systems may still produce high $z_{\rm GR}$ values.
For the second PCA parameter $\delta\hat{\phi}^{(2)}_{\rm PCA}$, GW151226-like injections generally give higher $z_{\rm GR}$ values than GW150914-like injections, except in the SNR 40 case of injection Nos. 1 and 5.
However, for $\delta\hat{\phi}^{(2)}_{\rm PCA}$, we find the deviations from GR are much less significant across the simulations, as mentioned earlier.
This makes the comparison across different systems less meaningful, as we have not quantified the statistical error bars on the $z$ scores, and the observed differences may, in many cases, lie within these unquantified uncertainties.
Broadly speaking, this trend indicates that the second PCA parameter is unable to detect GR violations for the specific types of fractional deviations used in this analysis, particularly for signals with ${\rm SNR}\le40$.

\subsection{FTI framework}
%%%%%%%%%%%%%%%%%%%%%%%%%%%%%%%%%%%%%%%%%%%%%%%%

\begingroup
\setlength{\tabcolsep}{0.75pt} % Default value: 6pt
\renewcommand{\arraystretch}{1.25} % Default value: 1
\begin{table*}[hbt!]
\begin{center}
\begin{tabular}{|c|c|c|c|c|c|c|c|c|c|c|}
\hline
\multicolumn{1}{| c |}{ No.} & \multicolumn{2}{c|}{Properties of Non-GR injections} & \multicolumn{4}{c|}{Properties of $\delta \hat{\phi}_{\rm PCA}^{(1)}$} & \multicolumn{4}{c|}{Properties of $\delta \hat{\phi}_{\rm PCA}^{(2)}$}\\
\cline{2-11}
\multirow{2}{*}{} & \multirow{2}{*}{$\{\delta\hat{\phi}_{k}\}$} & \multicolumn{1}{c|}{} & \multicolumn{2}{c|}{Median \&} & \multicolumn{2}{c|}{$\mathcal{Q}_{\rm GR}$} & \multicolumn{2}{c|}{Median \&} & \multicolumn{2}{c|}{$\mathcal{Q}_{\rm GR}$}\\
& & \multicolumn{1}{c|}{SNR} & \multicolumn{2}{c|}{90\% errors} & \multicolumn{2}{c|}{($z_{\rm GR}$)} & \multicolumn{2}{c|}{90\% errors} & \multicolumn{2}{c|}{($z_{\rm GR}$)}\\
\cline{4-11}
& & & GW1509 & GW1512 & GW1509 & GW1512 & GW1509 & GW1512 & GW1509 & GW1512\\
& & & 14-like & 26-like & 14-like & 26-like & 14-like & 26-like & 14-like & 26-like\\
\hline
1. & $\delta\hat{\phi}_{a}=0\, (a=-2,0,2)$, &  &  &  &  &  &  &  &  & \\
& $\delta\hat{\phi}_{b}=0.1\, (b=3,4,5l,$  & 20 & $0.11^{+0.11}_{-0.09}$ & $0.04^{+0.14}_{-0.12}$ & 0.02 (1.76) & 0.27 (0.47) & $-0.05^{+0.55}_{-0.60}$ & $-0.24^{+0.44}_{-0.59}$ & 0.56 (0.14) & 0.81 (0.76)\\
& $6,6l,7)$ &  &  &  &  &  &  &  &  & \\
\hline
2. & $\delta\hat{\phi}_{a}=0\, (a=-2,0,2)$, &  &  &  &  &  &  &  &  & \\
& $\delta\hat{\phi}_{b}=0.5\, (b=3,4,5l,$  & 20 & $0.51^{+0.11}_{-0.10}$ & $0.46^{+0.13}_{-0.14}$ & 0.00 (7.51) & 0.00 (5.28) & $0.11^{+0.62}_{-0.64}$ & $-0.03^{+0.42}_{-0.55}$ & 0.39 (0.29) & 0.54 (0.09)\\
& $6,6l,7)$ &  &  &  &  &  &  &  &  & \\
\hline
\hline
\end{tabular}
\caption{Summary of results from the simulated non-GR injection study within the {\tt FTI} framework that demonstrates the capability of the {\tt PCA-TGR} to detect GR violations in the PN phasing coefficients. The injected values of the fractional PN deviation parameter, injected binary systems, and the network SNRs (rounded to the nearest integer) of the injected signals are listed. For each non-GR injection, the table shows the median value, as well as the upper and lower limits of the 90\% credibility interval, $\mathcal{Q}_{\rm GR}$, and $z_{\rm GR}$ for the posteriors of $\delta\hat{\phi}^{(1)}_{\rm PCA}$ and $\delta \hat{\phi}^{(2)}_{\rm PCA}$. The posteriors on $\delta\hat{\phi}^{(1)}_{\rm PCA}$ exclude GR at $>3\,\sigma$ in the second case but not in the first case, while the posteriors on $\delta\hat{\phi}^{(2)}_{\rm PCA}$ encompass the GR value within 90\% credibility in both cases. The {\tt PCA-TGR} can rule out GR at $>3\,\sigma$ for a fractional deviation of magnitude $>0.5$ in the PN phasing coefficients.
}
\label{tab:table_non-GR_inj_fti}
\end{center}
\end{table*}
\endgroup
%%%%%%%%%%%%%%%%%%%%%%%%%%%%%%%%%%%%%%%%%%%

%%%%%%%%%%%%%%%%%%%%%%%%%%%%%%%%%%%
\begin{figure}[hbt!]
\centering
\includegraphics[scale=0.55] {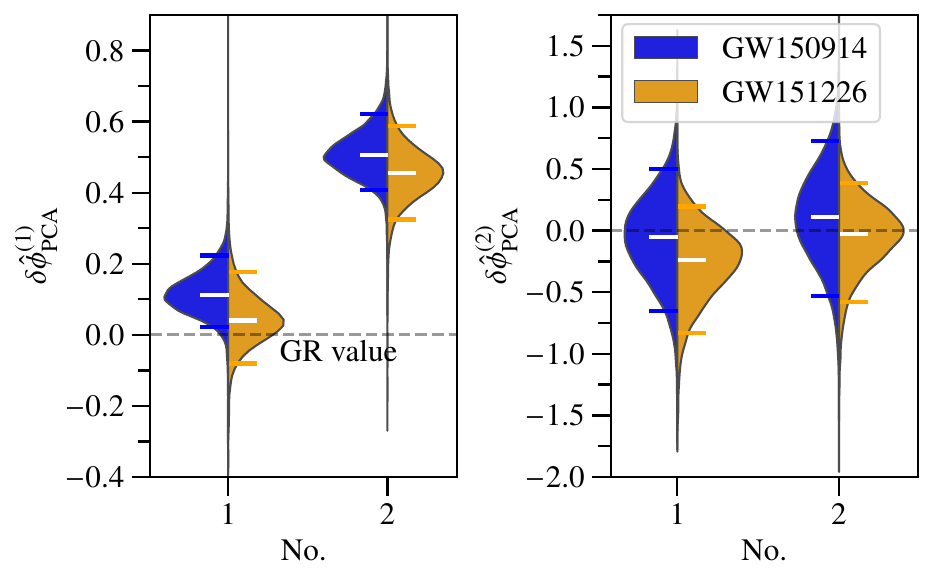}
    \caption{
Posteriors of $\delta\hat{\phi}^{(1)}_{\rm PCA}$ (left panel) and $\delta \hat{\phi}^{(2)}_{\rm PCA}$ (right panel) for each simulated injection listed in Table~\ref{tab:table_non-GR_inj_fti} presented as violin plots. The results from GW150914-like (GW151226-like) BBH injections are shown in blue (orange). 
The horizontal bars have the same meaning as in Fig.~\ref{fig:gr_inj_aligned_spin}. We mark the GR value of zero with dashed gray lines.
    }\label{fig:non-GR_fti}
\end{figure}
%%%%%%%%%%%%%%%%%%%%%%%%%%%%%%%%%%%

We now examine a couple of non-GR injections previously considered in the {\tt TIGER} framework but now analyzed within the {\tt FTI} framework. Specifically, we consider the first two cases in Table~\ref{tab:table_non-GR_inj}, which correspond to fractional deviations of 0.1 (where the {\tt PCA-TGR} method could not rule out GR at $>3\,\sigma$ in the {\tt TIGER} framework) and 0.5 (where the {\tt PCA-TGR} method could rule out GR at $>3\,\sigma$ in the {\tt TIGER} framework). The injection setups are identical to those in the previous subsection, except that here we use the {\tt parametrized SEOBNRv4HM\_ROM} waveform model both to generate the injected GW signals with GR violations (which therefore do not include the effects of precession) and to perform parameter estimation within the {\tt FTI} framework. We also restrict our analysis to scenarios with a network SNR of approximately 20.

The results from this analysis are presented in Fig.~\ref{fig:non-GR_fti} and Table~\ref{tab:table_non-GR_inj_fti}. Figure~\ref{fig:non-GR_fti} shows violin plots of the posterior probability distributions for the leading two PCA parameters from our injection analysis using the {\tt FTI} framework. The posterior medians, 90\% credible intervals, $\mathcal{Q}_{\rm GR}$, and $z_{\rm GR}$ values for the two leading PCA parameters for each injection are summarized in Table~\ref{tab:table_non-GR_inj_fti}. 
The posteriors on $\delta\hat{\phi}^{(1)}_{\rm PCA}$ exclude GR at $>3\, \sigma$ in the second case (corresponding to a fractional deviation of 0.5) but not in the first case (with a fractional deviation of 0.1). In contrast, the posteriors on $\delta \hat{\phi}^{(2)}_{\rm PCA}$ encompass the GR value within the 90\% credible interval in both cases. These results demonstrate that the {\tt PCA-TGR}, within the {\tt FTI} framework, can robustly identify GR violations at the $>3\, \sigma$ level when the deviation in the PN phasing coefficients exceeds 0.5 in magnitude—consistent with our earlier findings using the {\tt TIGER} framework.
However, in the case with a fractional deviation of 0.5, the posteriors on $\delta\hat{\phi}^{(1)}_{\rm PCA}$ in the {\tt FTI} framework show a slightly stronger deviation from GR compared to the {\tt TIGER} framework (see the $z_{\rm GR}$ values in Tables~\ref{tab:table_non-GR_inj} and~\ref{tab:table_non-GR_inj_fti}). This enhancement arises because the PCA parameters are slightly better constrained in the {\tt FTI} framework than in the {\tt TIGER} framework, due to differences in how the end of the inspiral regime is defined in the two frameworks (see the first paragraph of Sec.~\ref{sec:results_o3}).

In summary, we find that the {\tt PCA-TGR} is successfully able to detect GR violations of magnitude $0.5$ with a high confidence level. The ability to detect a GR violation depends on the magnitude of violation and the SNR of the signal.

\section{Response of PCA-TGR to Injections with Eccentricity}\label{sec:ecc-inj}

%%%%%%%%%%%%%%%%%%%%%%%%%%%%%%%%%%%%%%%%%%%%%%
%%%%%%%%%%%%%%%%%%%%%%%%%%%%%%%%%%%%%%%%%%%
\begingroup
\setlength{\tabcolsep}{0.75pt} % Default value: 6pt
\renewcommand{\arraystretch}{1.25} % Default value: 1
\begin{table*}[hbt!]
\begin{center}
\begin{tabular}{|c|c|c|c|c|c|c|c|c|c|}
\hline
\multicolumn{1}{| c |}{ID} & \multicolumn{3}{c|}{Properties of injections} & \multicolumn{2}{c|}{Properties of $\delta \hat{\phi}_{\rm PCA}^{(1)}$} & \multicolumn{2}{c|}{Properties of $\delta \hat{\phi}_{\rm PCA}^{(2)}$} & \multicolumn{2}{c|}{Properties of $\delta \hat{\phi}_{\rm PCA}^{(3)}$}\\
\cline{2-10}
\multirow{2}{*}{} & \multicolumn{1}{c|}{Mass ratio} & \multicolumn{1}{c|}{$e_{17}$} & \multicolumn{1}{c|}{Mode} & \multicolumn{1}{c|}{Median and} & \multicolumn{1}{c|}{$\mathcal{Q}_{\rm GR}$} & \multicolumn{1}{c|}{Median and} & \multicolumn{1}{c|}{$\mathcal{Q}_{\rm GR}$} & \multicolumn{1}{c|}{Median and} & \multicolumn{1}{c|}{$\mathcal{Q}_{\rm GR}$}\\
& & &  & \multicolumn{1}{c|}{90\% errors} & \multicolumn{1}{c|}{($z_{\rm GR}$)} & \multicolumn{1}{c|}{90\% errors} & \multicolumn{1}{c|}{($z_{\rm GR}$)} & \multicolumn{1}{c|}{90\% errors} & \multicolumn{1}{c|}{($z_{\rm GR}$)}\\
\hline
SXS:BBH:1155 & 1 & $<10^{-4}$ 
& (2,2)+(3,2) & $0.01^{+0.01}_{-0.02}$ & 0.27 (0.57) & $0.00^{+0.17}_{-0.19}$ & 0.48 (0.04) & $1.26^{+9.02}_{-10.33}$ & 0.41 (0.22)\\
\hline
SXS:BBH:1355 & 1 & 0.053 
& (2,2)+(3,2) & $-0.04^{+0.02}_{-0.03}$ & 1.00 (2.44) & $-0.54^{+0.86}_{-0.47}$ & 0.85 (1.33) & $1.59^{+12.27}_{-12.08}$& 0.42 (0.22)\\
\hline
SXS:BBH:1357 & 1 & 0.097 
& (2,2)+(3,2) & $0.25^{+0.05}_{-0.06}$ & 0.00 (7.83) & $-1.30^{+0.63}_{-0.86}$ & 1.00 (2.91) & $9.26^{+11.76}_{-22.08}$ & 0.25 (0.88)\\
\hline
SXS:BBH:1222 & 2 & $<10^{-4}$  
& (2,2)+(3,2) & $0.01^{+0.03}_{-0.04}$ & 0.37 (0.26) & $-0.01^{+0.28}_{-0.27}$ & 0.54 (0.08) & $-0.68^{+10.59}_{-10.94}$ & 0.54 (0.10)\\
\hline
SXS:BBH:1364 & 2 & 0.044 
& (2,2)+(3,2) & $0.07^{+0.04}_{-0.04}$ & 0.00 (2.75) & $-0.08^{+0.21}_{-0.21}$ & 0.73 (0.59) & $-14.25^{+8.67}_{-4.41}$ & 0.99 (3.49)\\
\hline
SXS:BBH:1368 & 2 & 0.097 
& (2,2)+(3,2) & $0.06^{+0.03}_{-0.02}$ & 0.00 (4.77) & $-0.64^{+0.25}_{-0.25}$ & 1.00 (4.20) & $-11.22^{+9.18}_{-9.16}$ & 0.98 (2.03)\\
\hline
SXS:BBH:2265 & 3 & $<10^{-4}$  
& (2,2)+(3,2) & $-0.08^{+0.22}_{-0.25}$ & 0.76 (0.52) & $0.80^{+1.70}_{-1.58}$ & 0.20 (0.82) & $-0.52^{+19.46}_{-18.93}$ & 0.52 (0.04)\\
\hline
SXS:BBH:1371 & 3 & 0.055 
& (2,2)+(3,2) & $-0.03^{+0.02}_{-0.03}$ & 0.99 (2.49) & $-0.18^{+0.16}_{-0.15}$ & 0.97 (1.87) & $-10.10^{+7.38}_{-6.07}$ & 0.98 (2.43)\\ 
\hline
SXS:BBH:1373 & 3 & 0.093 
& (2,2)+(3,2) & $0.28^{+0.04}_{-0.03}$ & 0.00 (13.22) & $-4.91^{+0.32}_{-0.39}$ & 1.00 (22.28) & $-3.53^{+18.34}_{-15.67}$ & 0.63 (0.34)\\
\hline
\hline
\end{tabular}
 \caption{Summary of results from the simulated eccentricity injection analysis performed in Sec.~\ref{sec:ecc-inj} that studies the response of the current implementation of the {\tt PCA-TGR} framework to the eccentric BBH signals. The properties of the simulated eccentricity injections are listed (see the text for more details). For each injection, the table shows the median value, as well as the upper and lower limits of the 90\% credibility interval, $\mathcal{Q}_{\rm GR}$, and $z_{\rm GR}$ for the posteriors of $\delta\hat{\phi}^{(1)}_{\rm PCA}$,  $\delta \hat{\phi}^{(2)}_{\rm PCA}$, and $\delta \hat{\phi}^{(3)}_{\rm PCA}$. The posteriors on PCA parameters in all the cases with $e_{17}<10^{-4}$ show consistency with GR at 90\% credibility. For the injections with $e_{17}\sim0.05$ and $e_{17}\sim0.1$, the posteriors on at least one of the PCA parameters exclude GR at  $>90\%$ credibility. 
 }\label{tab:table_ecc_sys}
\end{center}
\end{table*}
\endgroup
%%%%%%%%%%%%%%%%%%%%%%%%%%%%%%%%%%%%%%%%%%%
%%%%%%%%%%%%%%%%%%%%%%%%%%%%%%%%%%%%%%%%%%%%%%

%%%%%%%%%%%%%%%%%%%%%%%%%%%%%%%%%%%%%%%%%%%%%%%%%
%%%%%%%%%%%%%%%%%%%%%%%%%%%%%%%%%%%%%%%%%%%%%%%%%%%%%%%%%%%%%%%%%%%%
\begin{figure*}[hbt!]
    \centering
\includegraphics[width=0.75\textwidth]{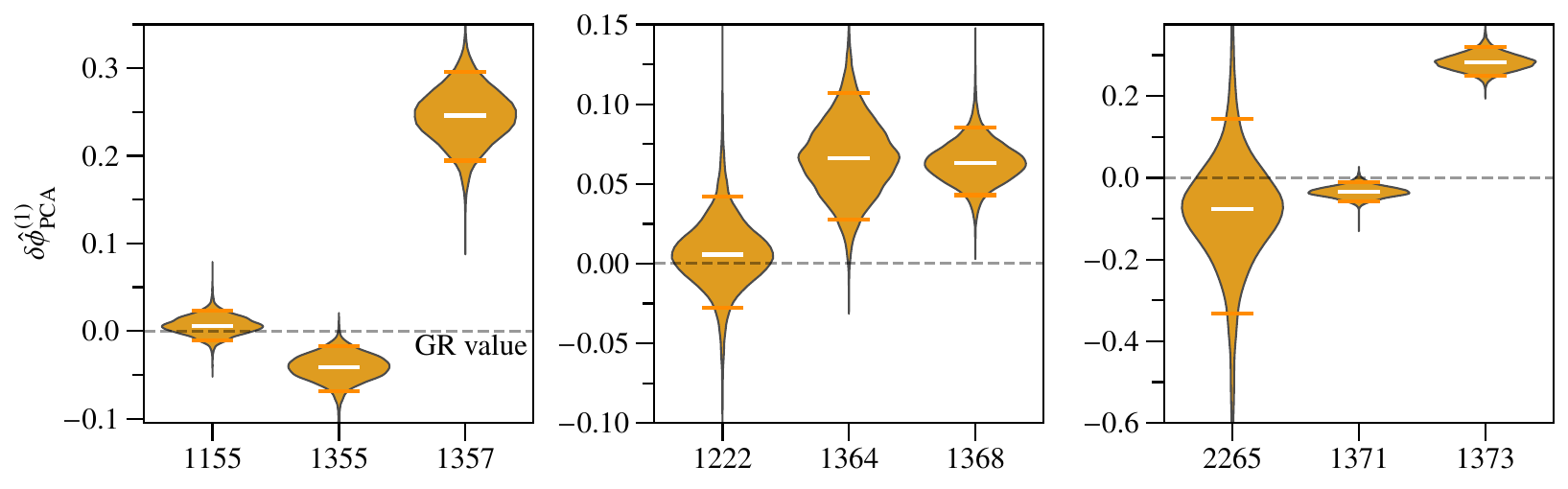}\\
\includegraphics[width=0.75\textwidth]{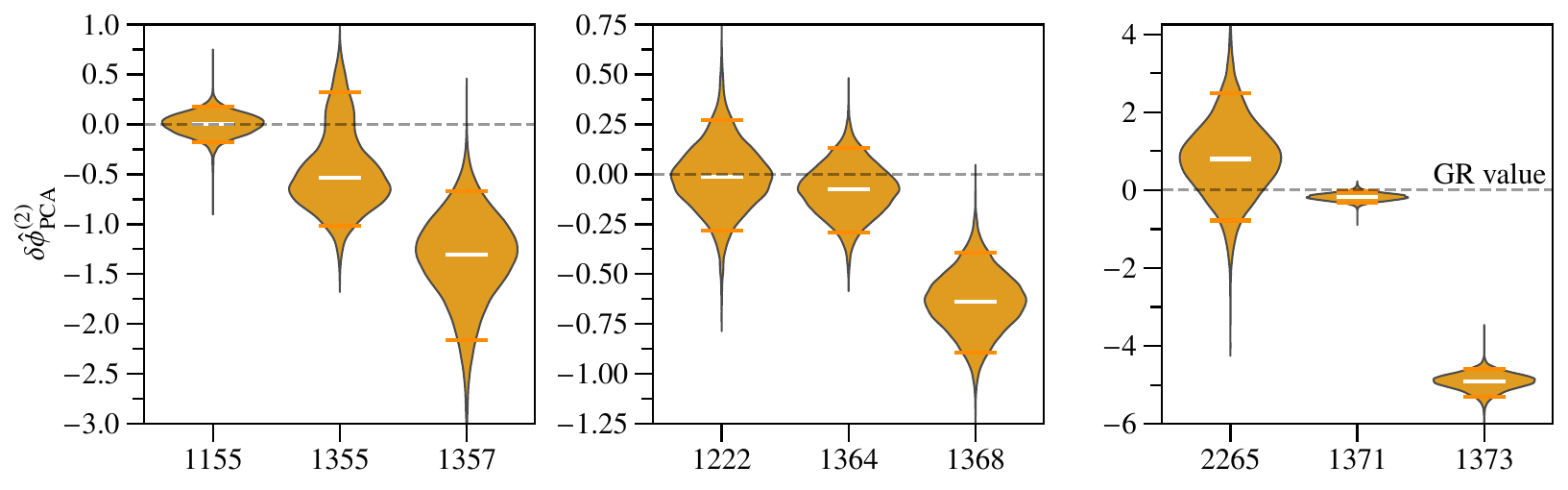}\\
\includegraphics[width=0.75\textwidth]{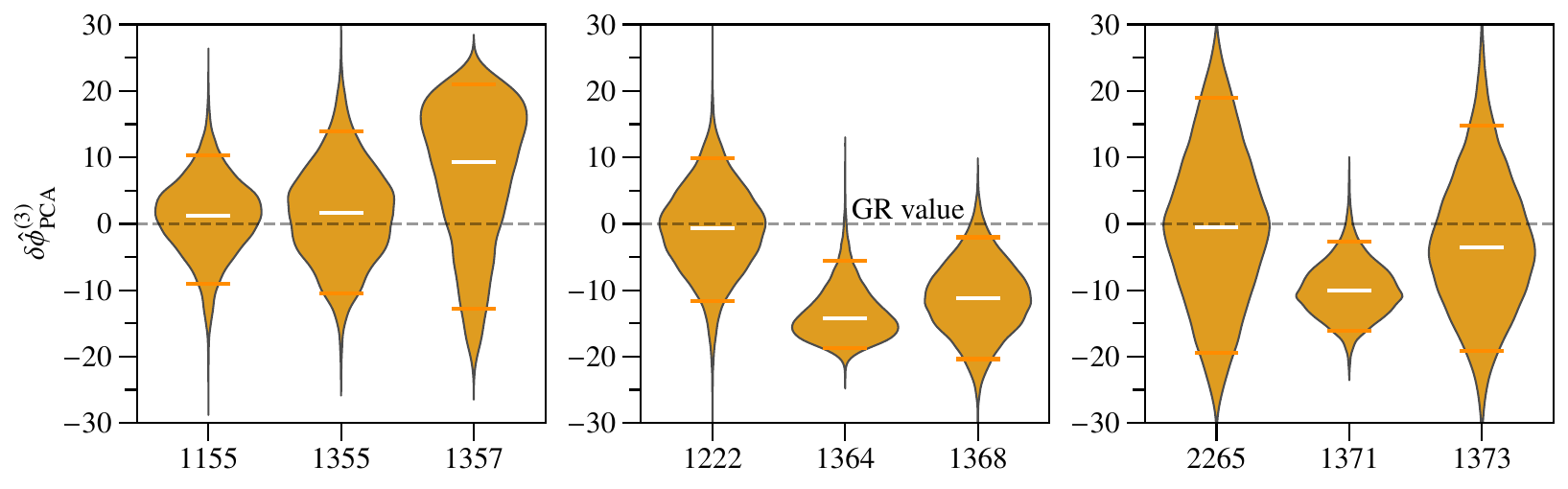}
    \caption{Posteriors of $\delta\hat{\phi}^{(1)}_{\rm PCA}$ (upper panel), $\delta \hat{\phi}^{(2)}_{\rm PCA}$ (middle panel), and $\delta \hat{\phi}^{(3)}_{\rm PCA}$ (lower panel) for each simulated eccentric injection listed in Table~\ref{tab:table_ecc_sys} presented as violin plots. The left, middle, and right panels show the results obtained from eccentric BBH injections with mass ratios 1, 2, and 3, respectively. The horizontal bars have the same meaning as in Fig.~\ref{fig:gr_inj_aligned_spin}. We mark the GR value of zero with dashed gray lines. 
    }
    \label{fig:ecc_sys}
\end{figure*}
%%%%%%%%%%%%%%%%%%%%%%%%%%%%%%%%%%%%%%%%%%%%%%%%%%%%%%%%%%%%%%%%
%%%%%%%%%%%%%%%%%%%%%%%%%%%%%%%%%%%%%%%%%%%%%%%%

Previous studies have found that the unmodeled eccentricity in the parametrized waveform models can lead to false violation of GR in {\tt TIGER}/{\tt FTI} tests if the BBH has significant residual eccentricity by the time it enters the detector frequency band~\cite{Saini:2022igm,Narayan:2023vhm}.  In this section, we study the response of {\tt PCA-TGR} to eccentric BBH signals when we employ a parametrized quasicircular inspiral-merger-ringdown waveform model for inference. The question is how the different PN deformation parameters of the quasicircular inspiral-merger-ringdown waveform respond to the neglect of eccentricity in the signal and how the response translates into the PCA parameters. 

We adopt the simulated eccentric BBH signals used in Ref.~\cite{Narayan:2023vhm}. These eccentric signals are nonspinning BBH NR simulations from the SXS catalog~\cite{Boyle:2019kee} (see also Ref.~\cite{Hinder:2017sxy}) with mass ratios $q=1, 2, 3$ and  a redshifted total mass of $80 M_{\odot}$ at a luminosity distance of 400 Mpc. Other extrinsic parameters take the values given in~\cite{Narayan:2023vhm}. For each mass ratio, we consider simulated signals with eccentricities, defined at $17\,{\rm Hz}$, $e_{17} \sim0.05$, $\sim0.1$ (representing binaries with small eccentricities) and $<10^{-4}$ (representative of BBH in quasicircular orbits).  
For each set of BBH parameters, we consider injections which have (2,2) and (3,2) modes of the NR simulation and analyze the simulated signal using the {\tt parametrized IMRPhenomXPHM} waveform model. The injection SNRs for the (2,2)+(3,2) mode for mass ratios $q=1,2,3$ are $67, 230$, and $137$, respectively. We briefly summarize the properties of the simulations adopted in Table~\ref{tab:table_ecc_sys}. The reader may refer to Sec.~III of \cite{Narayan:2023vhm} for a more detailed discussion of the simulations used here. 

The JS divergence between the posterior and prior distributions exceeds 0.1 bits for the first three leading PCA parameters in all injections, except for those with $(q=1, e_{17} = 0.053, 0.097)$ and $(q=3, e_{17} = 0.093, < 10^{-4})$, where the JS divergence for the third leading PCA parameter falls below 0.1 bits. Nevertheless, we include a discussion of the third PCA parameter $\delta\hat{\phi}^{(3)}_{\rm PCA}$, as it can carry a significant imprint of apparent GR violation in certain cases, as discussed below.
We show the violin plots for the posterior probability distributions of the leading three PCA parameters for all the eccentric injections in Fig.~\ref{fig:ecc_sys}. We also present the posterior median and 90\% credible intervals,  $\mathcal{Q}_{\rm GR}$, and $z_{\rm GR}$ of the leading three PCA parameters for each injection in Table~\ref{tab:table_ecc_sys}. As expected, the posteriors on PCA parameters in all the quasicircular ($e_{17}<10^{-4}$) injection analyses show consistency with GR at 90\% credibility. In all other eccentricity ($e_{17}\sim0.05$ and $e_{17}\sim0.1$) injection analyses, the posteriors on at least one of the PCA parameters (more than one PCA parameter in most of the cases) exclude GR at  $>90\%$ credibility. It can be seen that $\delta\hat{\phi}^{(3)}_{\rm PCA}$ is the least informative with respect to detecting a GR violation for most of the cases. For all the higher-eccentricity ($e_{17}\sim0.1$) injection cases, the posterior distributions on at least one of the PCA parameters show strong GR violation with GR excluded at the $>3\, \sigma$ level. More precisely, as expected, the amount of GR violation detected increases as eccentricity increases for all mass ratios. 
For $q=1,2$, $\delta\hat{\phi}^{(1)}_{\rm PCA}$ picks larger GR violation than $\delta\hat{\phi}^{(2)}_{\rm PCA}$, while for $q=3$ this trend is reversed and the second PCA parameter shows larger GR violation.
In the previous section, we saw that only the posteriors on $\delta\hat{\phi}^{(1)}_{\rm PCA}$ show strong GR deviation when one considers fractional deviations in PN coefficients from GR. However, in the eccentric injection analyses, in many cases the posteriors on $\delta\hat{\phi}^{(2)}_{\rm PCA}$ (and occasionally even the $\delta\hat{\phi}^{(3)}_{\rm PCA}$) also exclude GR at $>90\%$ credibility. 

Unlike the ``linear" fractional deviations we considered in the previous section, the eccentric BBH signal in GR induces ``nonlinear" modifications to the PN phasing of the waveform of BBH in a quasicircular orbit in GR. By nonlinear we mean deviations that cannot be simply absorbed into deformations of the existing PN phasing coefficients. Hence, when we perform a parameter estimation with {\tt parametrized IMRPhenomXPHM} on these injections, the eccentricity-induced corrections will show up as deviations in all of the PN coefficients with different degrees of departures which propagate into the posteriors on the PCA parameters as we see. Therefore, it is reasonable to expect that for generic departures from GR (or unknown waveform systematics)—not limited to the specific types of fractional deviations considered in the previous section—more than one PCA parameter can show statistically significant deviations from GR.

To summarize, we find that the unmodeled eccentricity in the parametrized waveform models might show up as an apparent GR violation in the posteriors of PCA parameters if we apply the {\tt PCA-TGR} method to the eccentric BBH signals. Furthermore, we find that the nonlinear modifications to the PN phasing could make the posteriors of the higher-order PCA parameters deviate significantly from GR.

\section{Analysis of GWTC-3 events}\label{sec:results_o3}
%%%%%%%%%%%%%%%%%%%%%%%%%%%%%%%%%%%%%%%%%%%%%%%%%
%%%%%%%%%%%%%%%%%%%%%%%%%%%%%%%%%%%%%%%%%%%%%%%%%%%%%%%%%%%%%%%%%%%%
\begingroup
\setlength{\tabcolsep}{0.75pt} % Default value: 6pt
\renewcommand{\arraystretch}{1.25} % Default value: 1
\begin{table}[b]
\begin{tabular}{|c|c|c|c|c|}
\hline
\multicolumn{1}{|c|}{\textbf{Events}} & 
\multicolumn{2}{c|}{\texttt{TIGER}} & 
\multicolumn{2}{c|}{\texttt{FTI}} \\
\cline{2-5}
& SNR & $f_{\rm max}$ & SNR & $f_{\rm max}$ \\
\hline
\hline
GW190412 & 17 & 127 & 20 & 257\\
\hline
GW190814 & 23 & 159 & 24 & 359\\
\hline
GW191204\_171526 & 17 & 260 & 18 & 522 \\
\hline
GW191216\_213338 & 16 & 126 & 19 & 322\\
\hline
\hline
\end{tabular}
\caption{Rounded-off optimal SNRs in the inspiral regime and the upper cutoff frequencies ($f_{\rm max}$), which define the end of the inspiral regime, of the four GWTC-3 events selected for the {\tt PCA-TGR} analysis using the {\tt TIGER} and {\tt FTI} frameworks.}
\label{tab:real_events}
\end{table}
\endgroup

%%%%%%%%%%%%%%%%%%%%%%%%%%%%%%%%%%%%%%%%%%%%%%%%%
%%%%%%%%%%%%%%%%%%%%%%%%%%%%%%%%%%%%%%%%%%%%%%%%%%%%%%%%%%%%%%%%%%%%

%%%%%%%%%%%%%%%%%%%%%%%%%%%%%%%%%%%%%%%%%%%%%%%%%
%%%%%%%%%%%%%%%%%%%%%%%%%%%%%%%%%%%%%%%%%%%%%%%%%%%%%%%%%%%%%%%%%%%%
\begin{figure}[hbt!]
    \centering
\includegraphics[width=0.475\textwidth]{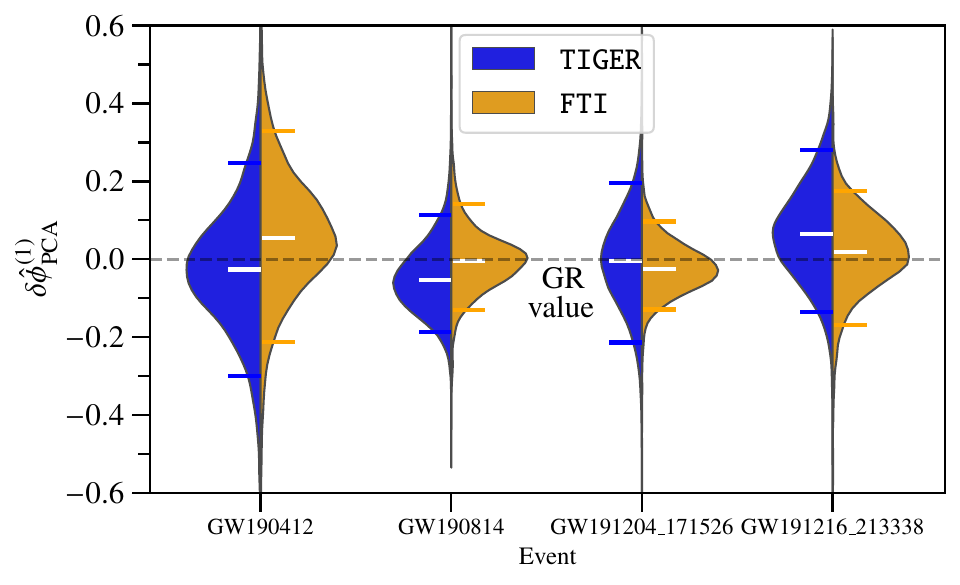}\\
\includegraphics[width=0.475\textwidth]{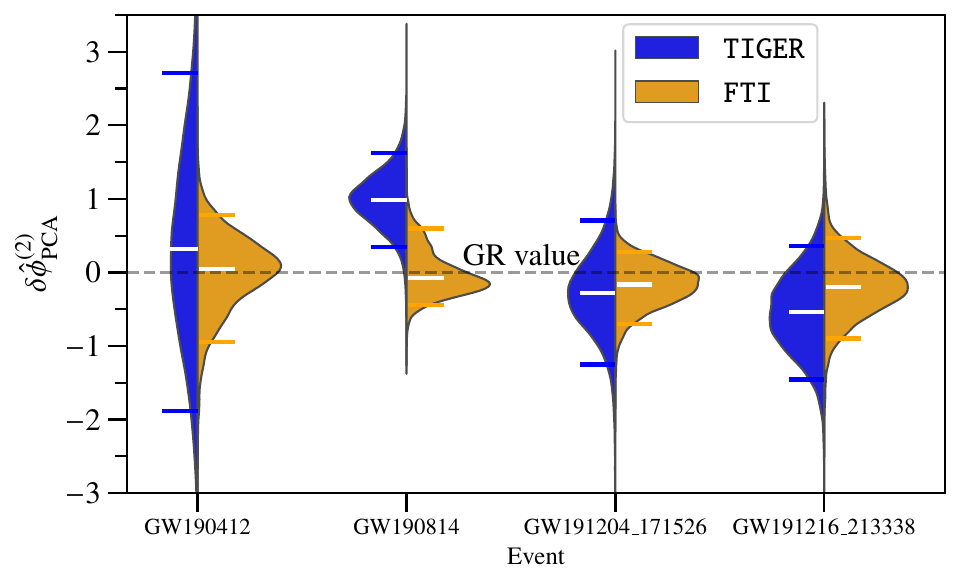}\\
\includegraphics[width=0.475\textwidth]{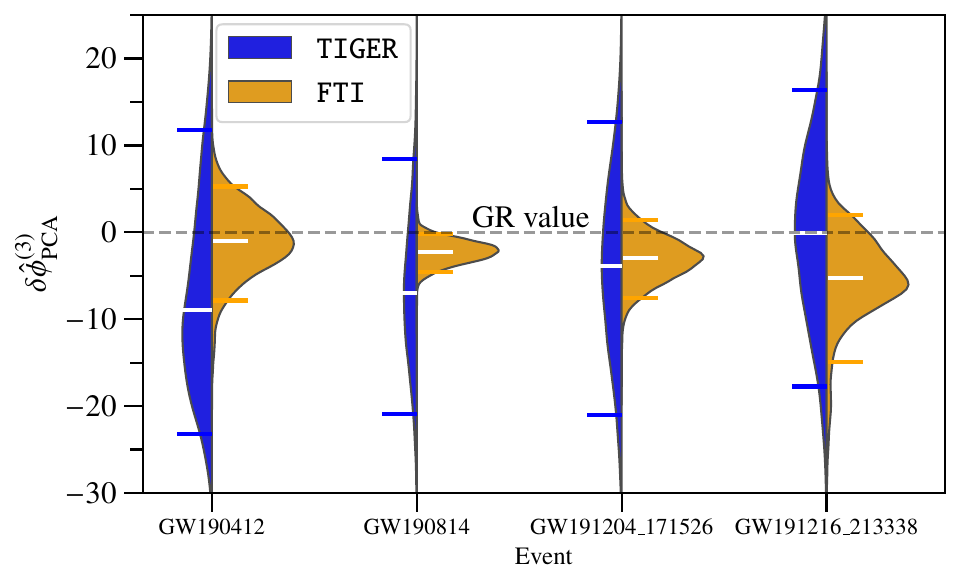}
\caption{ Posterior probability distribution of the first three leading PCA parameters from the GW events from GWTC-3 passing the selection criteria described in the first paragraph of Sec.~\ref{sec:results_o3} presented as violin plots. In each violin plot, the colored horizontal bars and the white solid line denote the 90\% credible intervals and the posterior median, respectively. We mark the GR value of zero with dashed gray lines.
    }
    \label{fig:pca_individuals}
\end{figure}
%%%%%%%%%%%%%%%%%%%%%%%%%%%%%%%%%%%%%%%%%%%%%%%%%%%%%%%%%%%%%%%%
%%%%%%%%%%%%%%%%%%%%%%%%%%%%%%%%%%%%%%%%%%%%%%%%

We now present the results obtained by applying the {\tt PCA-TGR} within the {\tt TIGER} and {\tt FTI} frameworks to selected GW events detected during the first three observing runs (O1, O2, and O3) of the Advanced LIGO and Advanced Virgo detectors~\cite{LIGOScientific:2016dsl,GWTC1,GWTC2,GWTC3}. The multiparameter runs involve sampling from 21- and 17-dimensional parameter spaces for precessing and aligned-spin BBHs, respectively, which is computationally expensive. Moreover, the combined bounds on the PCA parameters are largely dominated by the loudest and longest-duration inspiral signals in the catalog, which contribute the majority of the overall information content.
To manage computational cost, ensure a longer inspiral phase, and minimize the impact of noise artifacts, we impose selection criteria that reduce the number of BBH events analyzed while retaining the most informative signals. Specifically, we focus on events that are sufficiently loud---thereby reducing potential biases due to noise---by requiring a false-alarm rate of $\le 10^{-3}\, {\rm yr}^{-1}$, a network SNR $\geq 15$ in the inspiral regime, and a redshifted chirp mass $\leq 20\, M_{\odot}$.

The end of the inspiral regime is defined by the MECO frequency in the {\tt TIGER} framework and by the peak frequency of the (2,2) mode in the {\tt FTI} framework. 
The optimal SNR in the inspiral regime is computed using the maximum likelihood parameters from standard BBH parameter estimation under the GR hypothesis: The maximum likelihood parameters are extracted from parameter estimation samples with the tag {\tt C01:IMRPhenomXPHM}, provided by the LVK Collaboration and accessible on Zenodo~\cite{GWTC3-zenodo}. To compute the inspiral SNR, we use the {\tt IMRPhenomXPHM} waveform model for the {\tt TIGER} analysis and {\tt SEOBNRv4HM$\_$ROM} waveform model for {\tt FTI} analysis. 
For the chirp mass, we use the median of its posterior distribution from the same GR parameter estimation analysis. Both the optimal SNR in the inspiral regime and the chirp mass are rounded to the nearest integer. We find that only four events in GWTC-3 satisfy the selection criteria: GW190412, GW190814, GW191204\_171526, and GW191216\_213338. The optimal SNRs in the inspiral regime, along with the upper frequency cutoffs that define the end of the inspiral regime for these four selected events, are listed in Table~\ref{tab:real_events}. 

We perform the {\tt PCA-TGR} analysis on the above-chosen events within the {\tt TIGER} and {\tt FTI} frameworks. The {\tt FTI} multiparameter runs are performed with the {\tt parametrized SEOBNRv5HM$\_$ROM} waveform~\cite{Pompili:2023tna}, while the {\tt TIGER} multiparameter runs are performed with {\tt parametrized IMRPhenomXPHM-SpinTaylor} waveform~\cite{Colleoni:2024knd}~\footnote{Note that in all the injection analyses presented in the previous section, we used the {\tt parametrized IMRPhenomX} or {\tt parametrized IMRPhenomX-MSA}-version and the {\tt parametrized SEOBNRv4$\_$ROM}-version within the {\tt TIGER} and {\tt FTI} frameworks, respectively. However, in this section, we have used the newer {\tt parametrized IMRPhenomX-SpinTaylor}-version and {\tt parametrized SEOBNRv5$\_$ROM}-version in the {\tt TIGER} and {\tt FTI} frameworks, respectively. These newer versions are slightly improved compared to the previous versions, so we do not expect any changes in the results or conclusions presented in the previous sections.}. Fig.~\ref{fig:pca_individuals} shows the posterior distributions of the first three dominant PCA parameters that capture the most information (JS divergence between the prior and posterior distributions greater than 0.1 bits).
Although there are few cases, especially for the {\tt TIGER} framework, where only the first two leading PCA parameters have a JS divergence greater than 0.1 bits, we present the posterior distributions of the three most dominant PCA parameters for both {\tt FTI} and {\tt TIGER} for broad comparison. The reason why the third-dominant PCA parameter for {\tt FTI} is more informative than {\tt TIGER} is because the SNR and number of cycles in the inspiral regime for {\tt FTI} are larger than {\tt TIGER}. As mentioned before, the inspiral cutoff frequency for {\tt FTI} is the peak frequency of the (2,2) mode, 
which is larger than the MECO frequency used for {\tt TIGER} as the inspiral cutoff frequency.
The longer inspiral for the {\tt FTI} analysis compared to {\tt TIGER} enables the measurement of more PCA parameters 
and provides overall tighter constraints, if not comparable, as can be seen in Fig.~\ref {fig:pca_individuals}.
At the same time, the PN phase in Eq.~(\ref{eq:PN_phase}) becomes increasingly inaccurate at higher frequencies. Hence, the differences between the {\tt FTI} and {\tt TIGER} results can be considered as an estimate of the typical systematics of inspiral tests, with {\tt TIGER} providing more conservative bounds.

The bounds on the leading PCA parameters for all four events are $\mid \delta\hat{\phi}_{\mathrm{PCA}}^{(1)}  \mid\leq0.5$ and are consistent with GR for both {\tt TIGER} and {\tt FTI}. 
Of the four events, GW190814 provides the best constraints, which has the highest inspiral SNR and is the most asymmetric system, leading to the largest number of GW cycles in the inspiral regime. 
The second-best constraint is provided by GW191204\_171526 and GW190412, which have comparable inspiral SNRs as shown in Table \ref{tab:real_events}. 

The second leading PCA parameters for GW190412, GW191204\_171526, and GW191216\_213338 are consistent with GR for {\tt TIGER} and {\tt FTI} within the 90\% credible interval. However, for GW190814, the median of the posterior distribution for {\tt TIGER} shows a significant offset from the GR value. For {\tt TIGER}, the GR value lies at 0.01 quantile, indicating a statistically significant violation from the GR value. The posterior distribution of the second leading PCA parameter for {\tt FTI} is statistically consistent with GR. 
The composition of the second leading PCA eigenvector shows significant contributions from higher-order PN deformation parameters at 2.5PN, 3PN, and 3.5PN, indicating that these terms play a relatively larger role in shaping the posterior distribution peak away from the GR value.
This is consistent with the results obtained by the LVK Collaboration~\cite{GWTC2-TGR}, which showed that the GR value falls near the tails of the posterior distribution of the PN deformation parameters.

The maximum likelihood values of all mass parameters obtained from the multiparameter {\tt TIGER/FTI} runs differ from those obtained through standard parameter estimation using BBH waveforms in GR. In Sec.~\ref{subsec:prec_HM_inj}, we performed a GW190814-like GR injection under the zero-noise approximation in the {\tt TIGER} framework, using the {\tt IMRPhenomXPHM} waveform for injection and the {\tt parametrized IMRPhenomXPHM} waveform for recovery.
The posterior distributions of the first two dominant PCA parameters from this analysis are shown in blue in the rightmost plots of Fig.~\ref{fig:GR_inj_GW_events}. 
We see that the GR value is recovered well, which demonstrates that {\tt PCA-TGR} works robustly in the parameter space spanned by GW190814-like events. 
This injection study confirms that, at least, correlations between different parameters are not responsible for the offset observed in the PCA posteriors. For example, Ref. \cite{Sanger:2024axs} discusses a case where a degeneracy between the 0PN deviation parameter and the chirp mass led to an apparent GR violation in the posterior of the 0PN fractional PN deviation parameter for GW230529 \cite{GW230529}.
Therefore, the offsets in the posterior distribution of the second leading PCA parameter away from zero could be due to waveform systematics or noise artifacts in the data.

The posterior distributions of the third PCA parameter for {\tt TIGER} are not particularly informative for all events except GW190814, which also shows a significant offset from the GR value. For {\tt FTI}, the third PCA parameter also shows deviation from GR, especially for GW190814, for which the GR value lies at 0.96 quantile. 
The corresponding eigenvector of the third dominant PCA parameter has a significant contribution from the higher order PN deformation parameters, with the 3.5PN deformation parameter contributing $\sim50$\%. 
To understand the underlying reason for the deviations from GR shown by the most dominant PCA parameters, one first needs to examine various sources of possible false indications of GR violation~\cite{Gupta:2024gun}, such as the impact of unmodeled physics in the current waveform model or noise artifacts in the data. A detailed investigation of such waveform systematics or noise artifacts is beyond the scope of this paper.

\subsection{Joint bounds on the PCA parameters from selected GWTC-3 events}
\label{subsec:joint_bounds}
%%%%%%%%%%%%%%%%%%%%%%%%%%%%%%%%%%%%%%%%%%%%%%%%%%%%%%%%%%%%%%%%%%%%
\begin{figure}[hbt!]
    \centering
\includegraphics[width=0.975\columnwidth]{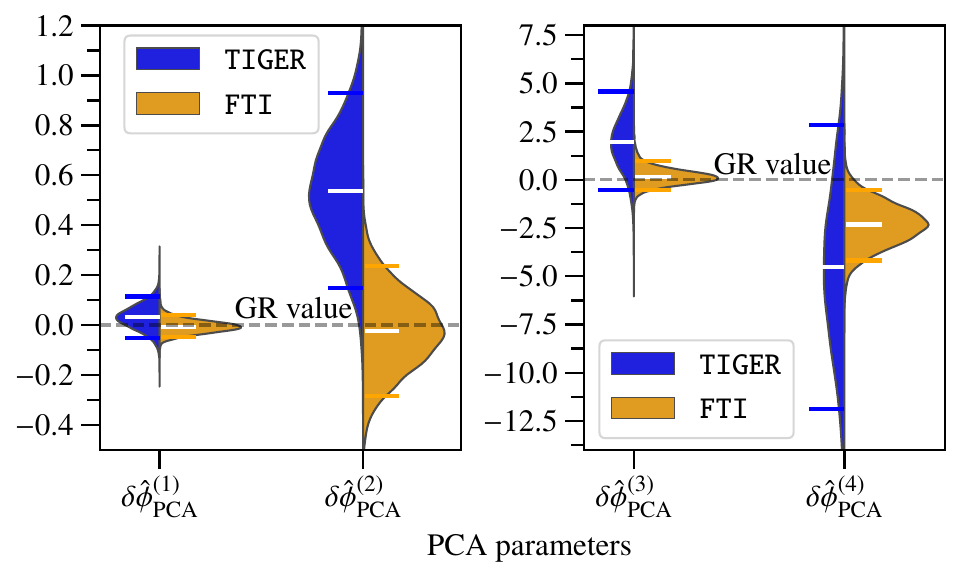}
    \caption{Joint posterior probability distributions on the first four leading PCA parameters from all the selected GWTC-3 events. The joint bounds are obtained through the marginalized likelihood multiplication technique. We mark the GR value of zero with dashed gray lines. 
    }
    \label{fig:pca_joints}
\end{figure}
%%%%%%%%%%%%%%%%%%%%%%%%%%%%%%%%%%%%%%%%%%%%%%%%%%%%%%%%%%%%%%%%

%%%%%%%%%%%%%%%%%%%%%%%%%%%%%%%%%%%%%%%%%%%%%%
%%%%%%%%%%%%%%%%%%%%%%%%%%%%%%%%%%%%%%%%%%%
\begingroup
\setlength{\tabcolsep}{0.75pt} % Default value: 6pt
\renewcommand{\arraystretch}{1.25} % Default value: 1
\begin{table}[t]
\begin{tabular}{|c|c|c|c|c|}
\hline
\multicolumn{1}{| c |}{$\delta\hat{\phi}_{\mathrm{PCA}}^{(i)}$} & \multicolumn{2}{c|}{Median and   90\% errors} & \multicolumn{2}{c|}{$\mathcal{Q}_{\rm GR}$ ($z_{\rm GR}$)}\\
\cline{2-5}
& {\tt TIGER} & {\tt FTI} & {\tt TIGER} & {\tt FTI}\\
\hline
\hline
$\delta\hat{\phi}_{\mathrm{PCA}}^{(1)}$ & $0.03^{+0.08}_{-0.08}$ & $-0.01^{+0.05}_{-0.04}$ & 0.27 (0.60) & 0.61 (0.26)\\
\hline
$\delta\hat{\phi}_{\mathrm{PCA}}^{(2)}$ & $0.54^{+0.39}_{-0.39}$ & $-0.02^{+0.26}_{-0.26}$ & 0.01 (2.26) & 0.56 (0.15)\\
\hline
$\delta\hat{\phi}_{\mathrm{PCA}}^{(3)}$ & $1.97^{+2.61}_{-2.49}$ & $0.14^{+0.82}_{-0.65}$ & 0.09 (1.24) & 0.36 (0.29)\\
\hline
$\delta\hat{\phi}_{\mathrm{PCA}}^{(4)}$ & $-4.52^{+7.36}_{-7.35}$ & $-2.32^{+1.80}_{-1.86}$ & 0.84 (1.01) & 0.98 ( 2.06)\\
\hline
\hline
\end{tabular}
\caption{Combined constraints on the leading four PCA parameters from all selected GWTC-3 events. The table shows the median, 90\% credible intervals, $\mathcal{Q}_{\rm GR}$, and $z_{\rm GR}$ of the combined PCA posteriors.
}
\label{tab:pca_combined}
\end{table}
\endgroup
%%%%%%%%%%%%%%%%%%%%%%%%%%%%%%%%%%%%%%%%%%%%%%
%%%%%%%%%%%%%%%%%%%%%%%%%%%%%%%%%%%%%%%%%%%

We follow the method described in Sec.~\ref{subsec:comb_events} to combine information from the selected GWTC-3 events, GW190412, GW190814, GW191204\_171526, and GW191216\_213338 to obtain joint bounds on the PCA parameters. We find that the joint posterior distribution of the first four most dominant PCA parameters, plotted in Fig.~\ref{fig:pca_joints}, have JS divergence greater than 0.1 bits, indicating that they carry significant information. The median and the 90\% credible interval for joint posteriors along with the corresponding GR quantiles are shown in Table~\ref{tab:pca_combined}.
For the joint posterior of the leading PCA parameter, the GR value lies at the 0.27 quantile in the {\tt TIGER} framework and the 0.61 quantile in the {\tt FTI} framework, indicating good consistency with the GR. 
However, the medians of the joint posterior distributions of the subdominant PCA parameters show a large offset from zero. In the {\tt TIGER} framework, the posterior distribution of the second PCA parameter exhibits the highest deviation from the GR value (GR at the 0.01 quantile), followed by the third (GR at the 0.09 quantile) and fourth (GR at 0.84 quantile). The trend is the opposite for {\tt FTI}, where the highest deviation from GR is shown by the least dominant fourth PCA parameter (GR at the 0.98 quantile), followed by the third (GR at the 0.36 quantile) and second (GR at the 0.56 quantile) PCA parameters.
We find the above interesting characteristics because of GW190814,\footnote{We have verified that the combined posteriors on the PCA parameters, using only three events, excluding GW190814, do not show any significant offset away from GR.} for which we get a statistically significant deviation from GR for the second PCA parameter in the case of {\tt TIGER} and a statistically significant deviation from GR for the third PCA parameter for {\tt FTI} as shown in  Fig. ~\ref{fig:pca_individuals}. The same is reflected in the joint posterior distributions of the subdominant PCA parameters but to a varied degree, as the joint posteriors are obtained after diagonalizing the joint posterior distribution of the original PN deformation parameters and not simply by multiplying the posteriors of a PCA parameter across events.

\section{Conclusions}\label{sec:conc}
It is likely that a deviation from GR will manifest as changes in multiple PN coefficients in the phase evolution of a compact binary. Therefore, a test that measures possible simultaneous departures from GR in these PN coefficients can act as a powerful test of GR. However, the high dimensionality of the parameter space—which includes both the standard GR parameters and fractional deviations in various PN coefficients, along with strong parameter degeneracies---poses a significant challenge to this test. Diagonalization of the covariance matrix that corresponds to the space of deformation parameters helps in identifying the best-measured linear combination of these parameters, and the relative weights of the new set of parameters helps to reduce the dimensionality of the parameter space. This variant of the multiparameter test is referred to as the principal component assisted test of GR or {\tt PCA-TGR}.

We have done an extensive study of the efficiency of {\tt PCA-TGR}, using the {\tt TIGER} and {\tt FTI} frameworks, that uses Phenom and EOB families of waveforms, respectively, to perform parametrized tests. Besides an extensive zero-noise injection campaign that covers aspects such as spin-precession and higher modes for GR as well as non-GR injections, we also studied how nonzero eccentricity would manifest as a (false) GR violation and how {\tt PCA-TGR} would respond to this for different values of eccentricities. We then applied this framework to some of the selected events in O3 that have long inspirals and found no statistically significant deviation from GR. We reported joint bounds by combining the constraints from these events.

Listed below are some of the caveats and future directions.
\begin{enumerate}
\item It is likely that any true deviation from GR in the inspiral phase will manifest as nonzero (and correlated) values of multiple PN deviation parameters $\delta\hat{\phi}_b$, which is the primary motivation for this PCA analysis. However, it is not possible to predict the pattern of these deviations for any generic beyond-GR model, or even (at present) for a specific beyond-GR theory. Additionally, it is not unlikely that any true deviation from GR in the inspiral phase will manifest as a different pattern of PN deviation parameters for different events (different masses, mass ratio, and spins). Thus, this analysis must be thought of as a test of the null hypothesis (GR is correct). Results that are inconsistent with zero deviation in the best-measured PCA combination(s) (both for the single-event analyses and especially for the multievent ``multiplication of likelihoods'' analysis) will require careful follow-up with other methods to understand the origin of the deviation. However, in the scenario that a true deviation from GR in the inspiral phase manifests as a fractional (small compared with 1, so approximately linear) deviation of a single PN parameter from zero (which may be the same or different parameters for different events), the PCA analysis has been demonstrated to be essentially as sensitive as the single parameter tests with {\tt TIGER} and {\tt FTI}.
\item Our results are based on 3.5PN phasing. However, when recently computed terms at 4 and 4.5PN are integrated into the {\tt TIGER}/{\tt FTI} frameworks, this method can be straightforwardly extended to include them.
 \item Currently, the multiparameter test on which PCA is performed simultaneously varies PN fractional deformation coefficients starting from 1.5PN onwards and does not include 0PN, 0.5PN, and 1PN orders. This is because the sensitivity of the current detectors is not good enough to break the degeneracies these PN order deformations have with chirp mass (0PN) and symmetric mass ratio (1PN). However, the algorithm allows this, and hence, if there is an event with high enough SNR and large enough inspiral cycles, such a test can be performed. 
 \item In its current implementation, {\tt PCA-TGR} deals with only fractional deformation parameters in the inspiral phase. Frameworks such as {\tt TIGER} allow for simultaneous deformations in the postinspiral phase of the compact binary waveform. This method can be straightforwardly extended to include such deformations, though a fresh injection campaign similar to the present one will be required to understand the results.
 \item Our mock non-GR waveforms are based on parametric waveforms and not from any specific theory of modified gravity. It would be interesting to use one of the inspiral-merger-ringdown waveforms from numerical simulations of modified gravity theories (such as Ref.~\cite{Okounkova:2019zjf}) and test how {\tt PCA-TGR} responds to such waveforms.
\end{enumerate}

Finally, we note that {\tt PCA-TGR} is not limited to parametrized PN tests—it can be applied to any parametrized null test of GR to derive constraints from multiparameter analyses. The framework is designed to be general and flexible, making it applicable to a wide range of theory-agnostic, beyond-GR null tests, such as BBH mimicker tests~\cite{Krishnendu:2017shb,Johnson-Mcdaniel:2018cdu}, parametrized ringdown tests~\cite{Carullo:2019flw,Maggio:2022hre,Pompili:2025cdc}, and more.

%TC:ignore
%%%%%%%%%%%%%%%%%%%%%%%%%%%%%%%%%%%%%%%%%
\section*{Acknowledgements}
\label{sec:acknowledgements}
We thank Anuradha Samajdar for critically reading the manuscript and providing useful comments. We are grateful to Nathan K. Johnson-McDaniel for his assistance with the NR eccentricity injection studies. P.M. thanks Alex~B.~Nielsen and N.~V.~Krishnendu for their insightful comments on the manuscript. P.M. thanks Duncan MacLeod for his suggestions on various computing issues. P.M. acknowledges the Science and Technology Facilities Council (STFC) for support through Grant No. ST/V005618/1. S.D. acknowledges support from the UVA Arts and Sciences Rising Scholars Fellowship. M.S. acknowledges support from the Weinberg Institute for Theoretical Physics at the University of Texas at Austin. K.G.A.~acknowledges support from the Core Research Grant No. CRG/2021/004565 of the Science and Engineering Research Board of India and a grant from the Infosys Foundation. K.G.A.~also acknowledges support from the Department of Science and Technology and the Science and Engineering Research Board of India via the Swarnajayanti Fellowship Grant No. DST/SJF/PSA-01/2017-18. P.D.R acknowledges support from the National Science Foundation (NSF) via NSF Award No. PHY-2409372. S.R. is supported by the Fonds de la Recherche Scientifique -- FNRS (Belgium). We also acknowledge NSF support via NSF Awards No. PHY-2308887 to P.N., No.~AST-2205920 and No.~PHY-2308887 to A.G., No.~PHY-2207780 to DS, No.~PHY-2309200 and No.~PHY-2207758 to A.J.W., and No.~AST-2307147, No.~PHY-2207638, No. PHY-2308886 and No.~PHY-2309064 to B.S.S. The authors are also grateful for computational resources provided by the Cardiff University and support by STFC Grants No.~ST/I006285/1 and No.~ST/V005618/1. The authors are grateful for computational resources provided by the LIGO Laboratory and supported by National Science Foundation Grants No. PHY-0757058 and No.
PHY-0823459. This manuscript has the LIGO preprint No. P2500459, and Weinberg Institute preprint No. UT-WI-25-2025.

This research has made use of data obtained from the Gravitational Wave Open Science Center (\url{https://www.gwosc.org/}), a service of the LIGO Laboratory, the LIGO Scientific Collaboration and the Virgo Collaboration. LIGO Laboratory and Advanced LIGO are funded by the United States National Science Foundation (NSF) as well as the Science and Technology Facilities Council (STFC) of the United Kingdom, the Max-Planck-Society (MPS), and the State of Niedersachsen/Germany for support of the construction of Advanced LIGO and construction and operation of the GEO600 detector. Additional support for Advanced LIGO was provided by the Australian Research Council. Virgo is funded, through the European Gravitational Observatory (EGO), by the French Centre National de Recherche Scientifique (CNRS), the Italian Istituto Nazionale di Fisica Nucleare (INFN) and the Dutch Nikhef, with contributions by institutions from Belgium, Germany, Greece, Hungary, Ireland, Japan, Monaco, Poland, Portugal, Spain. The construction and operation of KAGRA are funded by Ministry of Education, Culture, Sports, Science and Technology (MEXT), and Japan Society for the Promotion of Science (JSPS), National Research Foundation (NRF) and Ministry of Science and ICT (MSIT) in Korea, Academia Sinica (AS) and the Ministry of Science and Technology (MoST) in Taiwan.

\section*{DATA AVAILABILITY}
The data that support the findings of this article are not publicly available. The data are available from the authors upon reasonable request.

\bibliography{ref_list}
%TC:endignore

\end{document}